\documentclass[pre,aps,preprint,showpacs]{revtex4-1}
\usepackage{revsymb4-1,latexsym,amssymb,amsmath,comment}
\usepackage{graphicx,psfrag,subfigure,stmaryrd}
\usepackage[ansinew]{inputenc}

\newcommand{\be}{\begin{equation}}
\newcommand{\ee}{\end{equation}}
\newcommand{\bel}[1]{\begin{equation}\label{#1}}
\newcommand{\bea}{\begin{eqnarray}}
\newcommand{\eea}{\end{eqnarray}}
\newcommand{\ba}{\begin{array}}
\newcommand{\ea}{\end{array}}
\newcommand{\bef}{\begin{figure}}
\newcommand{\ef}{\end{figure}}

\begin{document}

\author{Thomas Bose and Steffen Trimper}
\affiliation{Institute of Physics,
Martin-Luther-University, D-06099 Halle, Germany}
\email{thomas.bose@physik.uni-halle.de}
\email{steffen.trimper@physik.uni-halle.de}
\title{Influence of randomness and retardation on the FMR-linewidth}
\date{\today }
\begin{abstract}
The theory predicts that the spin-wave lifetime $\tau_L$ and the linewidth of ferromagnetic
resonance $\Delta B$ can be governed by random fields and spatial memory. To that aim the effective field around
which the magnetic moments perform a precession is superimposed by a
stochastic time dependent magnetic field with finite correlation time. The magnetization
dynamics is altered by inclusion of a spatial memory effect monitoring a non-local interaction of size $\xi$.  The
underlying Landau-Lifshitz-Gilbert equation (LLG) is modified accordingly. The stochastic LLG is equivalent to
a Fokker-Planck equation which enables to calculate the mean values of the magnetization vector. Within the
spin-wave approximation we present an analytical solution for the excitation energy and its damping. The
lifetime and the linewidth are analyzed depending on the strength of the random field $D$ and its correlation
time $\tau_c$ as well as the retardation strength $\Gamma_0$ and the size $\xi$. Whereas $\tau_L$ decreases
with increasing $D$, retardation strength $\Gamma_0$ and $\tau_c$, the lifetime is enhanced for growing
width $\xi$ of the spatial retardation kernel. In the same manner we calculate the experimentally measurable linewidth
$\Delta B$ is increased strongly when the correlation time $\tau_c$ ranges in the nanosecond interval.
\end{abstract}

\maketitle   

\section{Introduction}

Ferromagnetic resonance (FMR) is a powerful technique to study magnetic materials, in particular the inherent magnetization
dynamics  \cite{HeinrichBland:Book:UMSIIandIII:2005,Hillebrands:Rezende:Mills:SDiCFII:2003}.
So the observable FMR-linewidth is very sensitive to the underlying dynamical processes as well as the real structure of the material like
anisotropy. From a theoretical point of view the Landau-Lifshitz-Gilbert equation (LLG) \cite{Landau:ZdS:8:p153:1935,Gilbert:ITOM:40:p3443:2004},
see Eq.~\eqref{LLG1} in the present paper, is an appropriate tool to investigate magnetic excitations and dissipative processes as the damping
of the excitations. Although, the LLG is known since a few decades it is still a standard model to analyze magnetodynamics. Recently the Gilbert damping 
parameter was determined experimentally for ferromagnetic thin films in \cite{Malinkowski:APL94:102501:2009} and by first-principle calculations for itinerant 
ferromagnets in \cite{Gilmore:PhysRevLett.99.027204:2007}. Obviously, the applicability of the LLG depends on the physical situation in mind. In case the magnetization 
is not conserved the Landau-Lifshitz-Bloch (LLB) equations are more appropriate, in particular in the vicinity of the phase transition as demonstrated in  
\cite{Chubykalo:PhysRevB74:Chubykalo:094436:2006}. The LLB equations were used to investigate magnetization switching near the Curie temperature 
in \cite{Garanin:PhysRevB70:212409:2004,Kazantseva:EPL86:27006:2009}. Moreover, the geometrical configuration of the sample are able to play an important role in measuring the 
FMR-linewidth. Related to this fact the contribution of the Gilbert damping to the linewidth can be superimposed by extrinsic effects as magnon-magnon scattering processes 
\cite{Sparks:PhysRev122:791:1961} which become of the same order of magnitude or even exceed the Gilbert damping. Especially for an in-plane configuration where the 
external field as well as the magnetization lie in the film plane the influence of two magnon processes to the FMR-linewidth cannot be neglected 
\cite{Arias:PhysRevB60:7395:1999,Arias:JAP87:5455:2000}. Those theoretical results predicting a nonlinear dependence of the linewidth on the frequency 
were extended to the case when the magnetization is tipped out of plane \cite{Landeros:PhysRevB77:214405:2008}. Different experimental findings 
emphasize the importance of extrinsic contributions for in-plane setups, 
see \cite{Lindner:PhysRevB68:060102:2003,Woltersdorf:PhysRevB69:184417:2004,Lenz:PhysRevB73:144424:2006,Zakeri:PRB76:2007:104416}. A quantitative separation of 
Gilbert damping and two magnon scattering contributions was carried out \cite{Lindner:PhysRevB68:060102:2003,Lenz:PhysRevB73:144424:2006,Zakeri:PRB76:2007:104416}.
Contrary to these observations there are other investigations \cite{Twisselmann:JAP93:6903:2003}, which offer no qualitative difference between in-plane 
and normal-to-plane measurements. In both realizations the linewidth depends linearly on the frequency even for frequencies smaller than $10~\rm{GHz}$ .
Such theoretical and experimental works suggest, among others, that the FMR-linewidth is exclusively controlled by the Gilbert damping and exhibits a pure linear frequency 
dependence in a perpendicular configuration with respect to thin films measurements. Furthermore, the two magnon scattering is supposed to be of less importance 
in bulk ferromagnets \cite{Lenz:PhysRevB73:144424:2006}. Thus the LLG equation seems still applicable to describe magnetization dynamics provided the physical situation 
is carefully analyzed as pointed out in \cite{Hillebrands:Rezende:Mills:SDiCFII:2003}. A more realistic magnetization dynamics requires a modification of the LLG. Recently, 
the anisotropic damping and its manifestation in the FMR-linewidth has been discussed by several authors 
\cite{FaehnleEtAl:JPhysD:0022-3727-41-16-164014,Seib:PhysRevB.79.092418,GilmoreStyles:PhysRevB.81.174414}. An alternative formulation of Gilbert damping by means of 
scattering theory was discussed in \cite{BraraasTserkovnyak:PhysRevLett.101.037207}.
In addition, ferromagnetic resonance measurements were used as well to investigate spin transport in magnetic single and double layer structures
\cite{KardaszHeinrich:PhysRevB.81.094409}.
Moreover, very recently it was shown that the transfer of spin angular momentum can induce ferromagnetic resonance dynamics in a
ferromagnetic film due to the spin Hall effect in an adjacent film with strong spin-orbit scattering \cite{LiuMoriyama:PhysRevLett.106.036601}.
Related to this phenomena it was reported on the direct time-resolved measurement of spin torque in magnetic tunnel junctions to detect
resonant magnetic precession due to an oscillating spin torque \cite{WangCui:Nat.Phys.2011:doi10.1038/nphys1928}.
A theory of ferromagnetic resonance in perpendicular magnetized nanodisks is suggested in \cite{AriasMills:PhysRevB.79.144404}.

To push forward the theory stochastic forces and non-local interactions should be included into the model to gain a more realistic description of magnetic materials
and to reveal unexpected behavior as for example the noise suppression by noise behavior argued in \cite{Vilar:RubiPhysRevLett.86.950}.
The effects of noise in magnetic nanosystems obeying spin torque dynamics are investigated in \cite{Foros:PRB:79:p214407:2009,Swiebodzinski:PhysRevB.82.144404}.
Experimentally, the role of noise in magnetic systems was prospected in \cite{DiaoNowak:PhysRevLett.104.047202,HartmannAPL96:2010:082108}.
The present work is addressed to the influence of randomness on the magnetization dynamics. As the two new aspects
the system considered is simultaneously subjected to feedback coupling and to a stochastic field with colored noise. The starting point is the LLG equation 
which is generalized in a manner that both spatial memory effects and a temporal stochastic field with a finite correlation time is
incorporated into the model. Previously the influence of colored noise \cite{BoseTrimper:PRB:81:104413:2010} and retardation
effects \cite{bosetrimper:retardatin:PRB:2011} within the LLG were analyzed separately. Otherwise, both effects can occur simultaneously.
Consequently we study a combined model concerning both kind of impacts, feedback and randomness. As demonstrated in former papers
there exits the possibility that the total damping, originated by the Gilbert damping and that one induced by memory effects are able to
cancel by the distinct damping mechanisms. In this paper we are interested in the FMR-linewidth. The corresponding parameters range in
such reasonable intervals where different dissipation sources are not observable. The main goal is to calculate the FMR-linewidth
and to discuss its dependence on the parameters characterizing randomness and retardation.

Let us give a brief outline of the paper. In Sec.~2 we present the mathematical model and its underlying basic assumptions. The stochastic LLG equation 
is equivalent to a Fokker-Planck equation which is derived approximately in Sec.~3. This equation enables to compute the mean values of
the magnetization. The results are discussed in detail in Sec.~4. Finally, we conclude by summarizing the results and by an outlook in Sec.~5.

\section{Model}

\begin{figure}[htb]%
  \includegraphics[width=.67\textwidth]{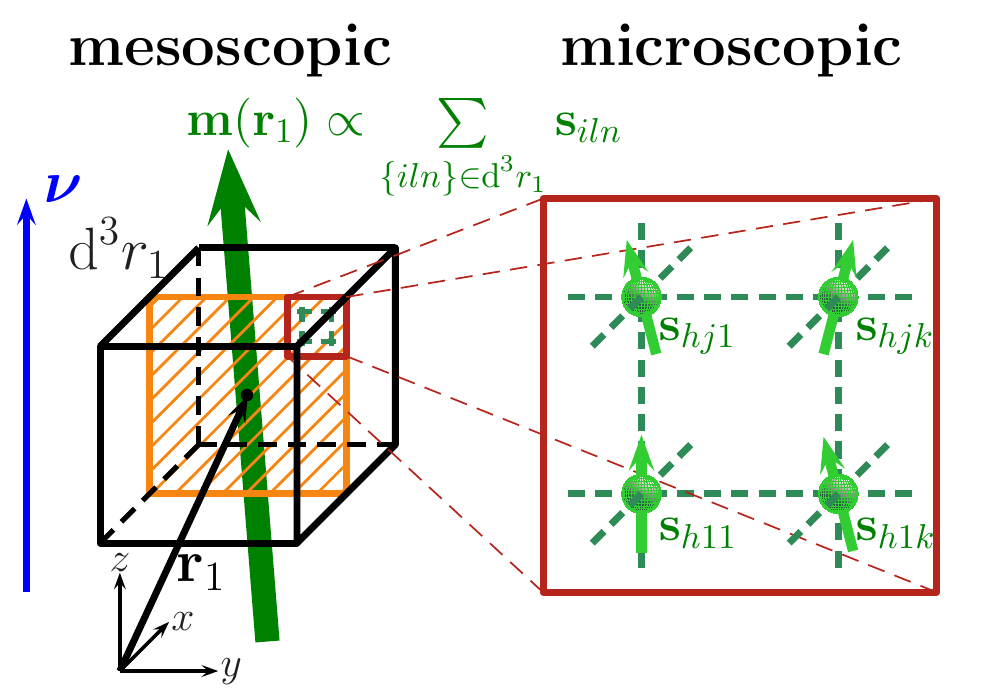}%
  \caption[]{%
    Illustration of the of the coarse-grained mesoscopic model. The $s_i$ represent microscopic magnetic moments which are related to the magnetization $\mathbf{m}$. Further explanation can be found in the text.}
    \label{picmeso}
\end{figure}

As already indicated in the introduction we are interested in micro- and nanosized magnets. Therefore a coarse-grained description is an appropriate tool
to investigate magnetic material. In this mesoscopic description the discrete magnetic moments are replaced by a spatiotemporal vector field $\bf{m}(\mathbf{r},t)$.
The interaction and the dynamics of the moments are formulated in a continuous approximation. The situation is schematically illustrated in Fig.~\ref{picmeso}.
Here, the magnetization $\mathbf{m}(\mathbf{r}_1)$ represents the magnetic properties within the volume-element $d^3r_1$ which is build around the position $\mathbf{r}_1$. The
field $\mathbf{m}(\mathbf{r}_1)$ stands for the total set of microscopic spins which will be visible if one zooms into the microscopic structure. The huge number of microscopic
degrees of freedom within $d^3r_1$ are substituted by a single degree of freedom, namely the mesoscopic quantity $\bf{m}$ which can be considered as the sum over the
microscopic spins located at equivalent crystal positions. Moreover, the magnetization vector field $\mathbf{m}(\mathbf{r}_1)$ is assumed to be oriented continuously in space.
The basics of our model consists in this mesoscopic description discussed before. Further, the system is supposed to offer an uniaxial anisotropy where the direction of the
anisotropy axis is denoted by $\boldsymbol\nu$. Our calculations refer to weak excitations which evolve as spin waves and possess a finite life time. Both quantities
are found in the long wave-length limit $qa\ll 1$, where $q$ is the amount of the wave vector and $a$ is the lattice constant. This assumption reflects the mesoscopic level
of description. Experimentally the dynamic behavior of the magnetization $\bf{m}$ can be detected for instance by means of ferromagnetic resonance (FMR). Because the main goal
of the paper is to put forward the modeling towards more realistic systems we develop a dynamic model for the magnetization field $\mathbf{m}(\mathbf{r})$ in which retardation effects
as well as stochastic fields are included. In particular, the aim is to relate our findings for the magnetic excitations and their damping to an experimentally
accessible quantity, namely the FMR-line width $\Delta B$, cf. Eq.~\eqref{LTandLW}.

As underlying model we start from is the Landau-Lifshitz-Gilbert equation
\be
\frac{\partial \mathbf{m}}{\partial t}=-\frac{\gamma}{1+\alpha ^2}\,\mathbf{m}\times \Bigl[ \mathbf{B_{\textrm{eff}}} +
\alpha \,[\bf{m}\times \mathbf{B_{\textrm{eff}}}]\Bigr] \,,
\label{LLG1}
\ee
which will be generalized accordingly. In Eq.~\eqref{LLG1} the quantities $\gamma$ and $\alpha$ are the gyromagnetic ratio and the dimensionless Gilbert
damping parameter, respectively. In this description $\mathbf{m}(\mathbf{r},t)$ is the unit vector \mbox{$\mathbf{m}=\mathbf{M}/M_s$} with the magnetization $\bf{M}$
the saturation magnetization. The local effective field $\mathbf{B_{\textrm{eff}}}(\mathbf{r},t)$ causes the precession of the magnetization. In general,
the effective field $\bf{B_{\rm{eff}}}$ is composed of different contributions, an internal field due to the interaction of the spins, the magnetic anisotropy 
and an external field. This effective field can be derived from the Hamiltonian of the system by functional variation with respect to $\mathbf{m}$
\be
\mathbf{B_{\textrm{eff}}} = - M_s^{-1} \frac{\delta \mathcal{H}}{\delta \mathbf{m}}\,.
\label{eff}
\ee
The Hamiltonian $\mathcal{H}$ can be expressed as \cite{BoseTrimper:PRB:81:104413:2010,Bar'Yakhtar:DynTopMagSol:Book:1994}
\begin{gather}
\begin{gathered}
\mathcal{H}=\int{d^3\mathbf{r}\,\{ w_{ex}+w_{an}+w_{ext}\}} \, \\
w_{ex}=\frac{1}{2}\,M_s\,\tilde{J} \,(\nabla \mathbf{m})^2 \,, \\
w_{an}=\frac{1}{2}\,M_s\,K \,\sin ^2\,\theta \quad , \quad w_{ext}=-\mathbf{B_{\textrm{ext}}}\cdot \mathbf{M} \,.
\label{fieldhamil}
\end{gathered}
\end{gather}
The quantities $\tilde{J} =J\,a^2$ and $K$ designate the exchange energy density and the
magneto-crystalline anisotropy energy density. Here, $J$ is the coupling strength between nearest neighbors referring to the isotropic Heisenberg model
\cite{Lakshmanan:PA:84:p577:1976} and $a$ is the lattice constant. Further, $\mathbf{B_{\textrm{ext}}}$ is the static external magnetic field. The quantity
$\theta$ is the angle between the the local magnetization $\mathbf{m}$ and the anisotropy axis \mbox{$\boldsymbol{\nu}=(0,0,1)$}. We assume that $\boldsymbol{\nu}$
points in the direction of the easy axis in the ground state. Therefore, $K>0$ characterizes the strength of the anisotropy. In deriving Eq.~\eqref{fieldhamil} we have
used $\mathbf{m}^2 = 1$. Let us stress again that this assumption seems to be correct if the temperature is well below the Curie temperature 
\cite{Chubykalo:PhysRevB74:Chubykalo:094436:2006}. Our calculations based on the LLG suggest that other damping mechanism such as an extrinsic magnon-magnon 
scattering due to magnetic inhomogeneities should be inactive and hence they are irrelevant. In thin films this situation can be achieved when both the 
magnetization and the static external field are perpendicular to the film plane. In 
our model this situation is realized when both the easy axis of the anisotropy $\boldsymbol\nu$ as well as the external field $\mathbf{B_{\textrm{ext}}}$ point into the 
$z$-direction. Hence the equilibrium magnetization is likewise oriented parallel to the $z$-axis. This situation corresponds to a normal-to-plane configuration. From here 
we conclude that the application of the LLG leads to reasonable results. For a different realization an alternative dynamical approach seems to be more accurate, 
see also the conclusions. 

\begin{figure}[htb]%
  \includegraphics[width=.67\textwidth]{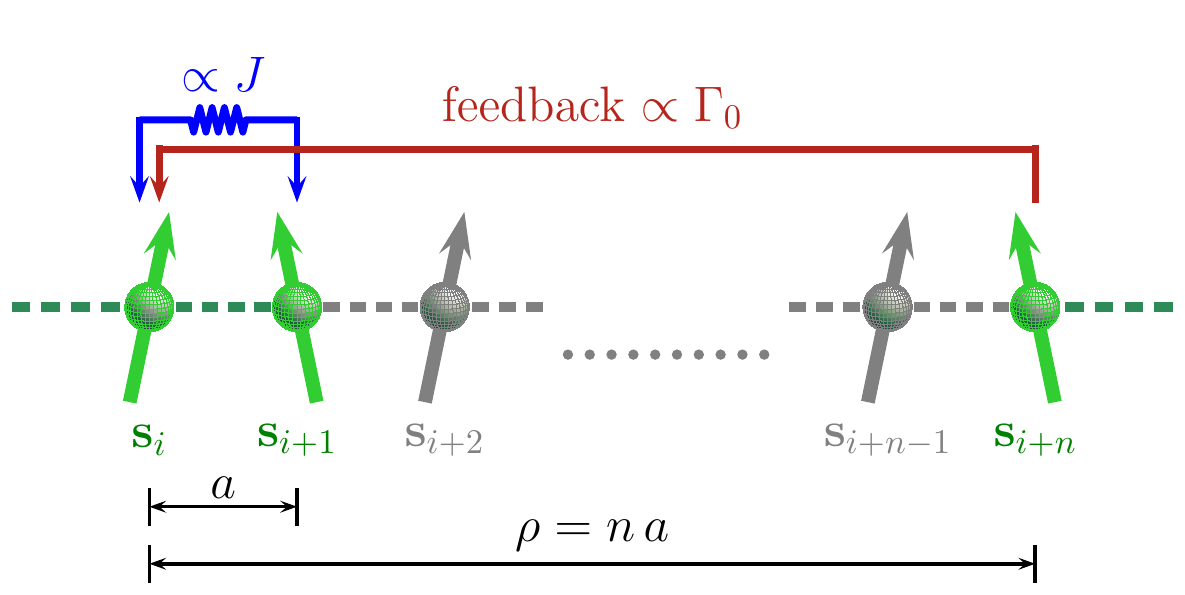}%
  \caption[]{%
    Schematic depiction of the difference between the exchange interaction $J$ and the coupling due to retardation $\propto \Gamma_0$. As is visible feedback mechanisms can
range over a larger distance $\xi \simeq \rho = n a$, where $n$ is integer.}
    \label{picret}
\end{figure}

To proceed further, the vector $\mathbf{m}$ is decomposed into a static and a dynamic part termed as $\boldsymbol{\mu}$ and
$\boldsymbol{\psi} = (\psi_1, \psi_2, \psi_3 )$, respectively.
In the frame of spin wave approximation we make the ansatz
\be
\mbox{$\mathbf{m}(\mathbf{r},t)$} =\boldsymbol\mu + \boldsymbol{\psi} (\mathbf{r},t) = \mu \,\boldsymbol{\nu}+\boldsymbol{\psi} \,,\quad \mu =
\textrm{const}.\,,
\label{ansatz}
\ee
Combining Eqs.~\eqref{eff} and \eqref{fieldhamil} yields the effective field
\be
\mathbf{B_{\textrm{eff}}}=\tilde{J}\,\nabla ^2\,\boldsymbol{\psi}-K\,\boldsymbol{\psi '} +\mathbf{B_{\textrm{ext}}} \quad ,
\quad \boldsymbol{\psi '} = (\psi _{1},\psi _{2},0) \,.
\label{Beff1}
\ee
It is appropriate to introduce dimensionless quantities:
\begin{gather}
\begin{gathered}
l_0^2 =\frac{\tilde{J}}{K}=\frac{J\,a^2}{K}\,,\quad \beta =(l_0\,q)^2 +1\,,\\
\Omega =\gamma \,K\,,\quad \bar{t}=\Omega \,t\,,\quad\frac{|\mathbf{B_{\textrm{ext}}}|}{K}=\varepsilon \,.
\label{rel1}
\end{gathered}
\end{gather}
The quantity $l_0$ is called the characteristic magnetic length \cite{Kosevich:PRSoPL:194:p117:1990} whereas the parameter $\varepsilon$ reflects the
ration between the strengths of the external and the anisotropy field. For convenience later we will substitute \mbox{$\bar{t}\rightarrow t$} again.
So far we have introduced the LLG in Eq.\eqref{LLG1} in its conventional form and incorporated our special basic model assumptions for a ferromagnetic material
below its Curie temperature. To proceed toward a more realistic description of magnets the LLG will be extended by the inclusion of retardation effects and random
magnetic fields. Whereas retardation is implemented by a memory kernel \mbox{$\Gamma (\mathbf{r},\mathbf{r'};t,t')$} a stochastic field $\boldsymbol\eta (\mathbf{r},t)$
contributes additionally to the effective field, i.e.
\be
\mathbf{B_{\textrm{eff}}}(\mathbf{r},t) \to \mathbf{b_{\textrm{eff}}}(\mathbf{r},t) =  \mathbf{B_{\textrm{eff}}}(\mathbf{r},t)+ \boldsymbol\eta (\mathbf{r},t)\,.
\label{stochas}
\ee
Taking both effects into account we propose the following generalized LLG
\begin{align}
\begin{split}
\frac{\partial \mathbf{m}(\mathbf{r},t)}{\partial t}=\int_0^t dt'\int d^d\mathbf{r'}\,\,
\Gamma (\mathbf{r}-\mathbf{r'};t-t')\, \\
 \times \biggl\{ -\frac{1}{1+\alpha ^2}\,\mathbf{m}(\mathbf{r'},t')\times \bigl[ \mathbf{b_{\textrm{eff}}}(\mathbf{r'},t')+ \bigr. \biggr. \\
\biggl. \bigl. +\alpha \,[\mathbf{m}(\mathbf{r'},t')\times \mathbf{b_{\textrm{eff}}}
(\mathbf{r'},t')] \bigr] \biggr\} \,,
\end{split}
\label{LLG3}
\end{align}
where the stochastic field is included in the dimensionless effective field as
\be
\mathbf{b_{\textrm{eff}}}=l_0^2\,\nabla ^2\,\boldsymbol{\psi}-\boldsymbol{\psi '}+\varepsilon \,\mathbf{b_0}+ \boldsymbol\eta (\mathbf{r},t) \,.
\label{Beff2}
\ee
The unit vector $\mathbf{b_0}$ indicates the direction of the external magnetic field.  In general, the kernel should respect the retardation concerning
temporal and spatial processes. More precise, a change of the magnetic moment at position $\bf{r}$ should influence another moment at position $\bf{r'}$ and vice versa.
This influence is thought to be an additional contribution which should not be confused with
parts of the exchange interaction in the effective field, i.g. the length $\xi$ on which spatial retardation effects are relevant could be of a different order of
magnitude in comparison with the lattice constant $a$. Insofar, a purely coordinate dependent part of the kernel reflects a kind of non-local interaction. All moments
within a radius $\xi$ contribute to the interaction. Likewise a temporal feedback mechanism can be taken into account due to the fact that the transport of information
from one magnetic moment to its neighbors needs at least a finite albeit small time. Such an in-time retardation mechanism is considered already in
\cite{bosetrimper:retardatin:PRB:2011}. Here we concentrate on instantaneous retardation in time whereas the spatial part is realized for simplicity
by a Gaussian shape
\be
\Gamma (\mathbf{r};t)=\delta (t)\,\left\{ \frac{\Gamma _0}{(\sqrt{\pi}\,\xi)^3}\,\exp\left[- \left( \frac{\mathbf{r}}{\xi}\right)^2 \right]\right\} \,,
\label{Gamma}
\ee
where $\Gamma_0$ and $\xi$ determine the strength and the size of the retardation, respectively. The
$\delta$-function in the last equation signalizes that all contribution to the interaction within
a sphere with radius $\xi$ contribute simultaneously to the interaction. As discussed below a typical value for
$\xi$ is assumed to be of the order $10^{-8} \rm{m}$, i.e. the time for the signal propagation within $\xi$ is about $10^{-15}-10^{-16} \rm{s}$.
Because this time is much smaller as the lifetime of the spin-waves, see the discussion below, we conclude that delay effects within the region with radius $\xi$
can be neglected. As indicated in Eq.~\eqref{stochas} the noise $\boldsymbol\eta (\mathbf{r},t)$ can also depend on space and time, i.e. in general random forces
can effect the value of the
magnetization at different positions in a distinct manner while additionally their fluctuations are also time dependent. Such a behavior maybe lead back to local
infinitesimal temperature gradients or defects. The random field $\boldsymbol\eta (\mathbf{r},t)$ is regarded as a colored noise the statistical properties of which
obey the following relations
\begin{align}
\begin{aligned}
\langle \eta_\alpha (t)\rangle =& 0 \,, \\
\chi _{\alpha \beta }(t,t') =& \langle \eta_\alpha (t)\,\eta_\beta (t')\rangle \\
 =& \frac{D_{\alpha \beta }}{\tau _{\alpha \beta }}\,\exp\left[-\frac{\mid t-t'\mid}{\tau _{\alpha \beta }}\right] \\
 &\xrightarrow[]{\tau _{\alpha \beta }\rightarrow 0} 2\,D_{\alpha \beta }\,\delta (t-t') \,.
\label{noise}
\end{aligned}
\end{align}
The components $\eta_\alpha (t)$ have a zero mean and a finite correlation time. As an aside in the limit $\tau \to 0$ the usual white noise properties are recovered.
However, we want to
concentrate on the more realistic colored noise case with $\tau > 0$. In Eq.~\eqref{noise} we assume $\eta_\alpha (\mathbf{r},t)=\eta_\alpha (t)$. In other words the total
system is affected by the same random influences. This may be reasonable if we have a well controllable constant temperature over the whole sample and an ideal sample without defects.
Let us briefly summarize the new properties of the model defined by Eqs.~\eqref{LLG3} and \eqref{Beff2}. The influences of retardation and a multiplicative noise as well
are implemented in the conventional Landau-Lifshitz-Gilbert equation. After a general incorporation into the model we had to limit the properties of both retardation and noise to
an idealized situation in order to obtain analytical results in the subsequent section. However, although each of the Eqs.~\eqref{Gamma} and \eqref{noise} represents a simplified
version of a more general case the linking between both by means of the equation of motion for the magnetization in Eq.~\eqref{LLG3} models a quite complex behavior which
is partly indicated in Fig.~\ref{picret}.
While the exchange interaction is a short range coupling over a lattice constant $a$, the interplay due to retardation with strength $\Gamma_0$ can cover a distance
$\rho$ which is a multiple of the lattice constant. If this distance $\rho$ is comparable to the characteristic length scale $\xi$ in Eq.~\eqref{Gamma} retardation effects
should be relevant. This microscopic picture can be transferred to a mesoscopic one and means a kind of non-local interaction. On the one hand at every spatial point the
same kind of noise affects the magnetization. Otherwise, the magnetization $\mathbf{m}(\mathbf{r},t)$ takes different values at distinct positions $\mathbf{r}$ and
therefore, the impact of the noise might be slightly different, too. Although spatial alterations of the noise are not regarded in the correlation function
defined in Eq.~\eqref{noise} the memory kernel respects spatial correlations within $\xi$ as seen in Eq.~\eqref{Gamma}. Insofar the effect of noise at different
spatial positions is transmitted by the memory kernel \mbox{$\Gamma (\mathbf{r},\mathbf{r'};t,t')$}. Another important hallmark is that the noise-noise correlation
function $\chi_{\alpha \beta }(t,t')$ is featured by a finite lifetime, cf. Eq.~\eqref{noise}. For the forthcoming calculations we assume that
$\tau_{\alpha \beta }=\tau_c\,\delta_{\alpha \beta }$. Likewise
the matrix of the noise correlation strength is supposed to be diagonal, i.e. $D_{kl}=D\,\delta_{kl}$. Hence, the two important stochastic parameters are the correlation
time $\tau_c$ and the correlation strength $D$ whereas the relevant parameters originated by the retardation are the retardation strength $\Gamma_0$ and the retardation
length $\xi$, see Eq.~\eqref{Gamma}. The results will be discussed in terms of the set of parameters $D, \tau_c, \Gamma_0$ and $\xi$.

\section{Statistical treatment}

Eqs.~\eqref{LLG3} and \eqref{Beff2} represents the stochastic LLG. Due to the coupling to the stochastic field $\boldsymbol\eta (\mathbf{r},t)$ the magnetization
field $\mathbf{m}(\mathbf{r},t)$ becomes a stochastic variable. To calculate the mean values of $\bf{m}$ one needs the probability distribution $P(\mathbf{m},t)$.
To that aim the current section is devoted to the derivation of an approximated Fokker-Planck equation which allows to  find the equations of motion for
averaged quantities. To that purpose let us reformulate the model presented in Eqs.~\eqref{ansatz}, \eqref{LLG3} and \eqref{Beff2}.
After Fourier transformation \mbox{$\varphi (\mathbf{q},t)=\mathcal{FT}\{ \psi (\mathbf{r},t)\}$} we find in linear spin-wave approximation
\be
\frac{d}{dt}\varphi _\alpha (\mathbf{q},t)=A_\alpha [\boldsymbol{\varphi}(\mathbf{q},t)]+\rm{B}_{\alpha \beta }[\boldsymbol{\varphi}(\mathbf{q},t)]\, \boldsymbol\eta_\beta (t)\,.
\label{sys}
\ee
The vector $\bf{A}$ and the matrix $\bf{B}$ posses the components
\be
\mathbf{A}=\frac{f(q,\xi)}{1+\alpha ^2} \begin{pmatrix}
										-(\beta \mu + \epsilon)\,(\alpha \mu \varphi_1 + \varphi_2) \\
											(\beta \mu + \epsilon)\,(\varphi_1 - \alpha \mu \varphi_2) \\
											0
									\end{pmatrix} \,\,,
\label{vectorA}
\ee
and
\be
\mathbf{B} =\frac{f(q,\xi)}{1+\alpha ^2} \begin{pmatrix}
						\alpha \mu \,\varphi _3 & \varphi _3 & -(\varphi _2+\alpha \mu \,\varphi _1) \\
						-\varphi _3 & \alpha \mu \,\varphi _3 & \varphi _1-\alpha \mu \,\varphi _2 \\
						\varphi _2 & -\varphi _1 & 0
					\end{pmatrix}\,.
\label{matrixB}
\ee
Here the function $f(\mathbf{q},\xi)$ is the Fourier transform of the memory kernel $\Gamma (\mathbf{r},t)$ defined in Eq.~\eqref{Gamma} and $\mu$ and $\beta$ are introduced in Eqs.~\eqref{ansatz} and \eqref{rel1}, respectively. Notice that $f$ depends only on the absolute value $q$ of the wave vector and takes the form
\be
f(q,\,\xi) = \Gamma_0\,\exp\left[ -\frac{1}{4}\,\xi^2 q^2\right] \,.
\label{fourierkernel}
\ee
To get the probability distribution function of the stochastic process determined by Eqs.~\eqref{noise} and \eqref{sys}-\eqref{matrixB} we define according to
\cite{Gardiner:HandbookStochMeth:Book:1990,Kampen:StochProcPhysChem:Book:1981}
\be
P(\boldsymbol{\varphi},t)=\left\langle \delta \,[\boldsymbol{\varphi}(t)-\boldsymbol{\varphi}] \right\rangle \,.
\label{probintro}
\ee
Here the symbol $<...>$ means the average over all realizations of the stochastic process. As usual $\boldsymbol\varphi (t)$ represents the stochastic process
whereas $\boldsymbol\varphi$ are the possible realizations of the process at time $t$. Due to the colored noise the corresponding Fokker-Planck equation
can be obtained only approximatively in lowest order of the correlation time. The time evolution of Eq.~\eqref{probintro} can be written in the form
\be
\frac{\partial}{\partial t}P(\boldsymbol{\varphi},t)=\mathcal{L}\,P(\boldsymbol{\varphi},t)\,.
\label{singleevent}
\ee
In deriving this expression we have used the time evolution of $\boldsymbol\varphi (t)$ according to Eq.~\eqref{sys}, the Novikov theorem \cite{Novikov:SPJ:20:p1290:1965}
and the correlation function given by Eq.~\eqref{noise} with $\tau_{\alpha \beta }=\tau_c\,\delta_{\alpha \beta }$, $D_{\alpha \beta }=D\,\delta_{\alpha \beta }$.
The form of the operator $\mathcal{L}$ is given in a correlation time and cumulant expansion while transient terms have been neglected
\cite{Fox:JoMP:18:p2331:1977,GarridoSancho:PhysA:1982:479,Dekker:PLA:90:p26:1982}
\begin{align}
\begin{aligned}
\mathcal{L}(\boldsymbol{\varphi},\tau_c )= &-\frac{\partial}{\partial \varphi _\alpha }A _\alpha (\boldsymbol{\varphi})+\frac{\partial}{\partial \varphi _\alpha }B _{\alpha \beta }
(\boldsymbol{\varphi})\frac{\partial}{\partial \varphi _\gamma } \\
&\times \Biggl\{ D\,\bigl[ B _{\gamma \beta }(\boldsymbol{\varphi})-\tau_c \,M_{\gamma \beta }(\boldsymbol{\varphi})\bigr] \Biggr. \\ &+
D^2\,\tau_c\,\left[ K_{\gamma \beta \mu  }(\boldsymbol{\varphi})\frac{\partial}{\partial \varphi _\nu }B _{\nu \mu }(\boldsymbol{\varphi}) \right. \\
&+\Biggl. \left. \frac{1}{2}B _{\gamma \mu }(\boldsymbol{\varphi})\frac{\partial}{\partial \varphi _\nu }K_{\nu \beta \mu }(\boldsymbol{\varphi})\right] \Biggr\} \,,
\label{Lnull}
\end{aligned}
\end{align}
with
\begin{align}
\begin{aligned}
M_{\gamma \beta } &= A _\nu \frac{\partial B _{\gamma \beta }}{\partial \varphi _\nu }-B _{\nu \beta }\frac{\partial A _\gamma }{\partial \varphi _\nu } \\
K_{\gamma \nu \beta } &= B _{\mu \beta }\frac{\partial B _{\gamma \mu }}{\partial \varphi _\mu }-\frac{\partial B _{\gamma \beta }}{\partial \varphi _\mu }B _{\mu \nu }\,.
\label{MKQ}
\end{aligned}
\end{align}
Notice that summation over double-indices is understood. The single probability distribution is determined by the operator $\mathcal{L}$ in Eq.~\eqref{Lnull} which
enables us to find the equation of motion for the expectation values $\left\langle \varphi _\alpha \right\rangle$. It follows
\begin{align}
\begin{aligned}
\frac{d}{dt}\left\langle \varphi _\alpha (t)\right\rangle =& \left\langle A _\alpha \right\rangle +D\,\left\langle \frac{\partial B _{\alpha \beta }}{\partial \varphi _\gamma }
\bigl( B _{\gamma \beta }-\tau_c \,M_{\gamma \beta } \bigr) \right\rangle \\
&-D^2\,\tau_c \,\Biggl\{ \left\langle \frac{\partial}{\partial \varphi _\nu } \left( \frac{\partial B _{\alpha \beta }}{\partial \varphi _\gamma }K_{\gamma \beta \kappa }\right)
B _{\mu \kappa }\right\rangle \Biggr. \\
														&+\Biggl. \frac{1}{2}\,\left\langle \frac{\partial}{\partial \varphi _\nu } \left( \frac{\partial B _{\alpha \beta }}
{\partial \varphi _\gamma }B _{\gamma \kappa }\right) K_{\mu \beta \kappa } \right\rangle \Biggr\}\,.
\label{EWphi}
\end{aligned}
\end{align}
Notice that in the white noise case all terms $\propto \tau_c$ would vanish.

\section{Results and discussion}

We find an analytical solution for the colored noise problem in Eq.~\eqref{EWphi} by standard Greens' function technique and Laplace transformation. After performing the
summation in Eq.~\eqref{EWphi} while making use of Eqs.~\eqref{vectorA}, \eqref{matrixB} and the expressions in Eq.~\eqref{MKQ} the result reads
\be
\left\langle \boldsymbol{\varphi}(t)\right\rangle =\begin{pmatrix}
						e^{-\delta \,t}\cos( \Omega \,t) & e^{-\delta \,t}\sin( \Omega \,t) & 0 \\
						-e^{-\delta \,t}\sin( \Omega \,t) & e^{-\delta \,t}\cos( \Omega \,t) & 0 \\
						0 & 0 & e^{-\sigma \,t}
					\end{pmatrix}\cdot \left\langle \boldsymbol{\varphi _0}\right\rangle \,,
\label{SOLphi}
\ee
where $\left\langle \boldsymbol{\varphi_0}\right\rangle =\left\langle \boldsymbol{\varphi}(t=0)\right\rangle$ are the initial conditions.
Physically, the parameters $\delta ,\,\sigma$ and $\Omega$ play the roles of the inverse magnon lifetimes and the frequency of the spin wave, respectively.
They are determined by
\begin{align}
\begin{aligned}
\delta =& \alpha \,\mu \,[\varepsilon +\beta \,\mu ]\,\frac{f(q,\xi)}{1+\alpha ^2} + D\,[2-\alpha^2\,\mu^2]\,\left( \frac{f(q,\xi)}{1+\alpha ^2}\right)^2 \\
&+ 2\,D\,\tau_c\,\alpha \,\mu \,[\varepsilon + \beta \,\mu]\,\left( \frac{f(q,\xi)}{1+\alpha ^2}\right)^3 \\
&+ D^2\,\tau_c \,[1-6\,\alpha^2\,\mu^2] \,\left( \frac{f(q,\xi)}{1+\alpha ^2}\right)^4 \,, \\
\Omega =& -[\varepsilon +\beta \,\mu ]\,\frac{f(q,\xi)}{1+\alpha ^2} + 3\,D\,\alpha \,\mu \,\left( \frac{f(q,\xi)}{1+\alpha ^2}\right)^2 \\
&+ D\,\tau_c\,[\alpha^2\,\mu^2 -1]\,[\varepsilon + \beta \,\mu]\,\left( \frac{f(q,\xi)}{1+\alpha ^2}\right)^3 \\
&+ \frac{D^2\,\tau_c}{2}\,\alpha \,\mu[11-3\,\alpha^2\,\mu^2]\,\left( \frac{f(q,\xi)}{1+\alpha ^2}\right)^4 \,, \\
\sigma =& 2\,D\,\frac{f(q,\xi)}{1+\alpha ^2}^2 -[4\,D\,\tau_c\,\alpha \,\mu \,(\varepsilon +\beta \,\mu )]\,\frac{f(q,\xi)}{1+\alpha ^2}^3 \\
&+ D^2\,\tau_c\,[3\,\alpha^2\mu^2 +1]\,\frac{f(q,\xi)}{1+\alpha ^2}^4 \,.
\label{delomsi}
\end{aligned}
\end{align}
Note that the parameters of the retardation mechanism, the strength $\Gamma_0$ and the length scale $\xi$, are included in the function $f(q,\xi)$ defined in
Eq.~\eqref{fourierkernel}. The two important parameters originated from the noise are the correlation time $\tau_c$ and the correlation strength $D$ of the random force.
Both affect the quantities in Eq.~\eqref{delomsi} as well. We proceed by studying the system under the variation of these four model parameters. To be comparable to
FMR experiments we refer to the following quantities
\begin{gather}
\begin{gathered}
\tau_L = (\delta \,\gamma \,K)^{-1} \,,\,\Delta B = 1.16 \, \frac{\alpha \,\omega}{\gamma} = 1.16\,\alpha \,K\,\Omega \,,
\label{LTandLW}
\end{gathered}
\end{gather}
i.e. the lifetime $\tau_L$ of the spin waves and the FMR-linewidth $\Delta B$, compare \cite{HeinrichBland:Book:UMSIIandIII:2005,Zakeri:PRB76:2007:104416}, which are related to
the dimensionless inverse lifetime $\delta$ and frequency $\Omega$ from Eq.~\eqref{delomsi}. Here the frequency $\omega$ is tantamount to the resonance frequency of the spin waves.
The lifetime $\tau_L$ and the linewidth $\Delta B$ are given in SI-units. Notice that the frequency independent part $\Delta B_0$, typically added on the right-hand side 
of the equation for $\Delta B$ is already subtracted in Eq.~\eqref{LTandLW}. The contribution $\Delta B_0$ is supposed to take into account magnetic inhomogeneities. 
For a quantitative evaluation we need to set the model parameters to reasonable values. In doing so
we also refer to Eq.~\eqref{rel1}. First let us start with fixed values. For the Gilbert damping parameter we choose the bulk value for Co which was found to be
$\alpha \simeq 0.005$ \cite{Tserkovnyak:PRL:88:p117601:2002,Katine:PhysRevLett:84:3149}. A similar value ($\alpha \simeq 0.0044$) was measured for a $\rm{FE}_4/\rm{V}_4$ 
multilayer sample in perpendicular configuration where only intrinsic Gilbert damping is operative \cite{Lenz:PhysRevB73:144424:2006}.

\begin{figure}[htb]%
  \includegraphics[width=.75\textwidth]{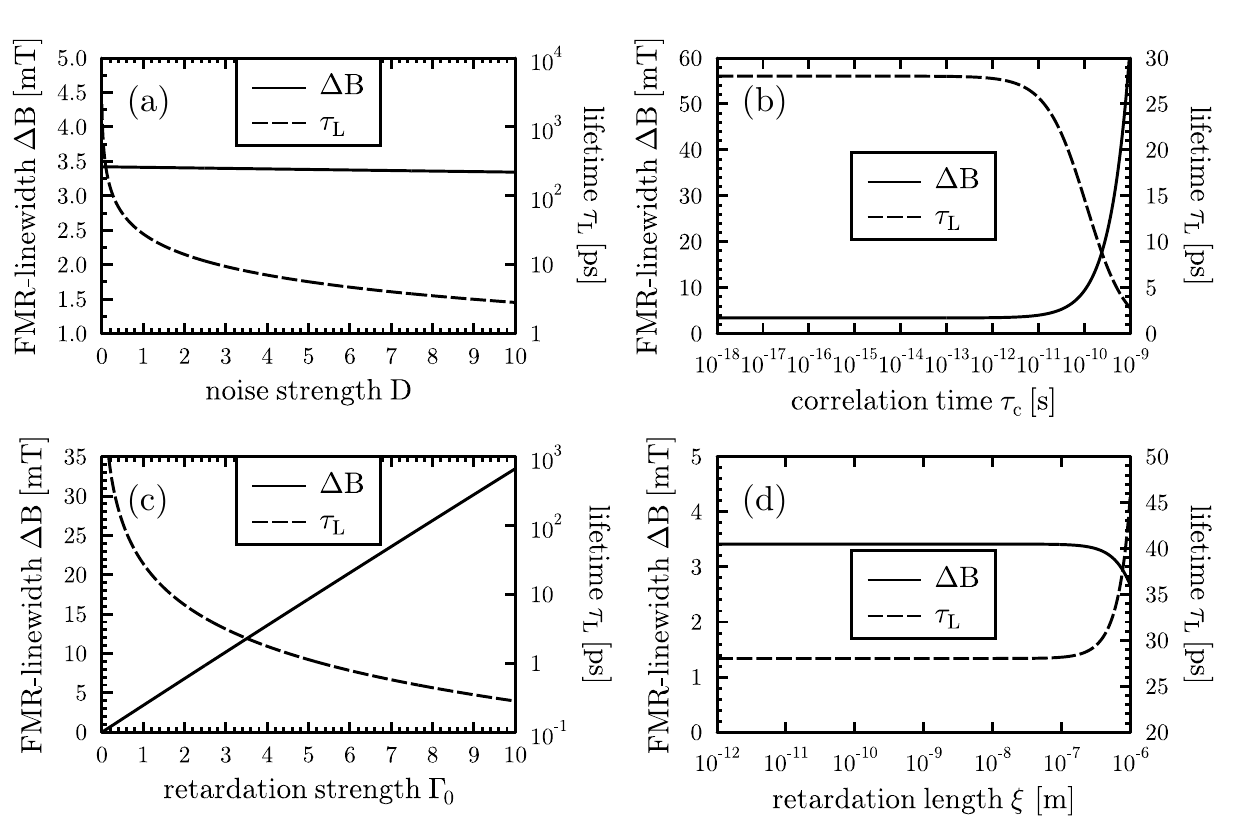}%
  \caption[]{%
   The FMR-linewidth and the lifetime depending on: (a) the noise correlation strength $D$ for $\tau_c=568\,\rm{as}$, $\Gamma_0=1$, $\xi =10^{-8}\,\rm{m}$; (b) the noise correlation time $\tau_c$ for $D=1$, $\Gamma_0=1$, $\xi =10^{-8}\,\rm{m}$; (c) the retardation strength $\Gamma_0$ for $\tau_c=568\,\rm{as}$, $D=1$, $\xi =10^{-8}\,\rm{m}$; (d) the retardation length $\xi$ for $\tau_c=568\,\rm{as}$, $\Gamma_0=1$, $\Gamma_0$. The other parameters take $l_0 =10^{-8}\,\rm{m}$, $q=10^6\,\rm{m}^{-1}$, $\varepsilon =5$, $\mu =0.9$ and $\alpha =0.005$. }
    \label{piclinewidth}
\end{figure}

\begin{figure}[htb]%
\includegraphics[width=\linewidth]{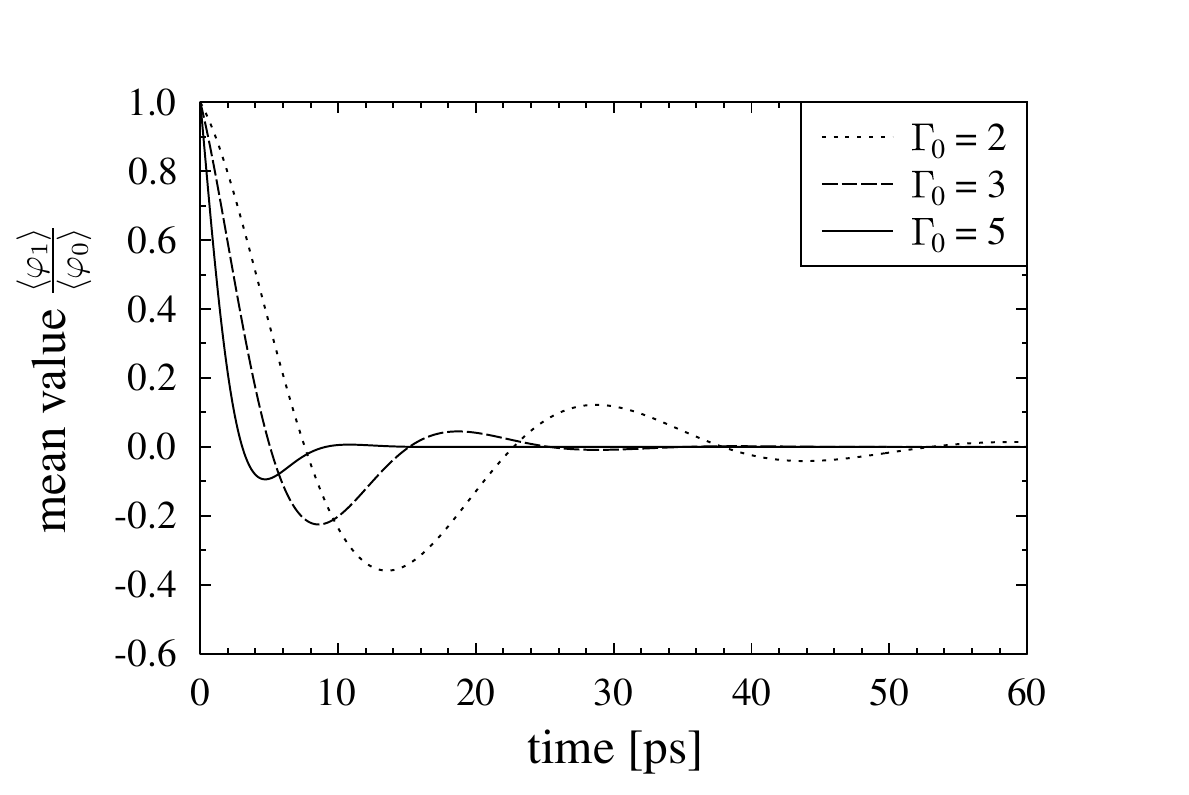}
\caption{%
  Evolution of spin waves for different values of the retardation strength $\Gamma_0$. The other parameters take $l_0 =10^{-8}\,\rm{m}$, $q=10^6\,\rm{m}^{-1}$, $\varepsilon =5$, $\mu =0.9$ and $\alpha =0.005$, $\tau_c=568\,\rm{as}$, $D=0.5$ and $\xi =10^{-8}\,\rm{m}$.}
\label{picspinwave}
\end{figure}

The anisotropy field $K$ is estimated as follows. Since the exchange interaction is
typically about $10^4$ times larger than relativistic interactions which are responsible for anisotropy \cite{Landau:StatPhysPart2:Book:1980} and the magnetic exchange field
can adopt large values we estimate the anisotropy as $K=0.1\,\rm{T}$. Since we are interested in small excitations transverse to the anisotropy axis $\boldsymbol\nu$
we suppose $\mu=0.9$ for the time independent part of the magnetization pointing in the direction of the anisotropy axis $\boldsymbol\nu$, compare Eq.~\eqref{ansatz}.
Moreover, the gyromagnetic ratio $\gamma \simeq 1.76\times 10^{11}\,(\rm{Ts})^{-1}$. The characteristic magnetic length defined in Eq.~\eqref{rel1} is of the order
$l_0 \simeq 10^{-8}\,\rm{m}$. For the calculations a static magnetic field of about $0.5\,\rm{T}$ is taken into account which corresponds to the scaled external
field $\varepsilon = 5$. Notice that the dispersion relation $\Omega$ in Eq.~\eqref{delomsi} is $q$-dependent. In the following we assume a medial value
$q=10^6\,\rm{m}^{-1}$. The parameters which are altered in the upcoming analysis are the noise correlation strength $D$ and the retardation strength $\Gamma_0$.
We investigate our model for both values ranging in the interval [0,10]. After this estimation the two parameters, the noise correlation time $\tau_c$ and the
retardation length $\xi$ are left over. For a comprehensive estimation we suggest that $\xi$ is ranged in $ 10^{-12}\,\rm{m} < \xi < 10^{-6}\,\rm{m}$. The lower limit is
smaller than a typical lattice constant $a \simeq 10^{-10}\,\rm{m}$ where the upper limit is a few orders of magnitude larger than the lattice constant. Likewise
the correlation time $\tau_c$ captures a quite large interval. Remark that we keep the notation $\tau_c$, especially with regard to Fig.~\ref{piclinewidth}, although we
also designated the dimensionless correlation time as $\tau_c$. In order to cover a wide range the time interval is chosen in between atto- and nanoseconds.
The results are depicted in Fig.~\ref{piclinewidth} and Fig.~\ref{picspinwave}.
In Fig.~\ref{piclinewidth} the behavior of the FMR-linewidth $\Delta B$ as well as the lifetime of the spin waves $\tau_L$, introduced in Eq.~\eqref{LTandLW}, are shown in
dependence on the different model parameters explained above. The influence of the correlation noise strength on $\Delta B$ and $\tau_L$ is shown in Fig.~\ref{piclinewidth}(a).
Whereas the linewidth decreases only very weak linearly when the noise strength $D$ is increased, the lifetime of the spin waves $\tau_L$ reveals a strong dependency on $D$.
This is indicated by the fact that $\tau_L$ is monotonic decaying while it covers several orders of magnitude with growing noise strength $D$.
The curve shape for the lifetime $\tau_L$ seems to be comprehensible because the stronger the stochastic forces are correlated and interact mutually the faster the coherent
motion of the spin moments is destroyed. This microscopic picture is reasonable under the premise that the evolution of spin waves is based on the phase coherence
between adjacent magnetic moments. Apparently the frequency and consequently the linewidth $\Delta B$ show only a quite small effect, compare Eq.~\eqref{LTandLW}.
Therefore, the variation of $D$ reveals no significant influence on the frequency velocity of the moments. A distinct behavior is depicted in Fig.~\ref{piclinewidth}(b)
for $\Delta B$ and $\tau_L$ as a function of the noise correlation time $\tau_c$. Both the linewidth and the lifetime remain constant for large interval of the correlation
time $\tau_c$, roughly speaking for $\tau_c$ ranging from $\rm{as}$ to $\rm{ps}$. If the correlation time is in between $\rm{ps}$ and $\rm{ns}$ the linewidth $\Delta B$
increases about a factor of $20$ and the lifetime $\tau_L$ decreases to a value about $9$-times smaller. Thus $\tau_c$ affects both $\Delta B$ and $\tau_L$ in an
opposite manner provided the noise-noise correlations occur on time scales larger than $\rm{ps}$. In this regime a growing correlation time $\tau_c$ implicates
likewise an enhancement of the resonance frequency of the spin waves $\omega \propto \Delta B$, see Eq.~\eqref{LTandLW}. Simultaneously the spin wave lifetime $\tau_L$
declines strongly. Such a behavior may be attributed to a 'stochastic acceleration' which on the one hand enhances the frequency but on the other hand
drives neighboring magnetic moments out of phase coherence. Remark that for times $\tau_c > 1\rm{ns}$ the linewidth $\Delta B$ tends to infinity. This effect is not shown in the
picture. Concerning the influence of the retardation parameters we refer to Fig.~\ref{piclinewidth}(c), which illustrates the influence of the retardation strength $\Gamma_0$.
As recognizable the FMR-linewidth exhibits a seemingly linear dependence as function of $\Gamma_0$ while $\Delta B$ grows with increasing retardation strength. The lifetime
$\tau_L$ decreases in a non-linear manner. The decay covers a range of $\approx 3$ orders of magnitude. We suggest the following mechanism behind this effect:
Let us consider two moments both localized at arbitrary positions within the retardation length $\xi$ as schematically displayed in Fig.~\ref{picret}. The mutual coupling
due to retardation between both characterized by $\Gamma_0$ leads to a phase shift between neighboring spins. Therefore, the phase coherence originated by the self-organized
internal magnetic field is interfered in view of an interplay within the feedback coupling in coordinate space. The stronger this interaction $\Gamma_0$ is the faster
is the damping of the spin waves. Accordingly,  spin wave solutions for different values of $\Gamma_0$ are plotted exemplary in Fig.~\ref{picspinwave}.
The retardation length $\xi$ influences $\Delta B$ and $\tau_L$ as well as is visible in Fig.~\ref{piclinewidth}(d). Here the quantities $\Delta B$ and $\tau_L$ remain
constant for a retardation strength $\xi$ ranging within the $\rm{pm}$ regime and a few tenth $\rm{\mu m}$.
For larger $\xi$-values the linewidth $\Delta B$ decreases while the lifetime $\tau_L$ increases. In the regime $\xi > 1 \rm{\mu m}$ the linewidth
$\Delta B$ tends to zero and the lifetime $\tau_L \to \infty$. This behavior is not depicted in Fig.~\ref{piclinewidth}(d). Notice that for reasonable values of $\xi$ which
not exceed the sample size the mentioned situation is not realized. The shapes of the curves in Fig.~\ref{piclinewidth}(d) may be explained as follows. This graph corresponds
to a fixed retardation strength $\Gamma_0$ while the retardation length $\xi$ is enlarged. Again we refer to the physical picture where the internal field, originated by the
mutual interaction of the moments, and the coupling due to the retardation operate as opposite mechanisms. The interplay happens in such a manner that
an increasing retardation strength $\Gamma_0$ weakens or destroys the phase coherence between adjacent spins.
Yet it is found that a growing retardation length $\xi$ counteracts the damping of the spin waves. As a consequence we suppose that the more spins are involved into
the retardation effect, i.e. the larger the parameter $\xi$ becomes, the more the damping is reduced. In other words it seems that retardation effects can
average out if sufficiently many magnetic moments are involved.

\section{Conclusions}

In the present paper we have studied a model on a mesoscopic scale realized by means of Landau-Lifshitz-Gilbert dynamics. The magnetization is driven by an
effective magnetic field. This field consists of an internal field due to the exchange interaction, an anisotropy field and a static external field. Additionally,
the effective field is supplemented by a time depending random one obeying colored noise statistics. Moreover, the stochastic LLG is generalized by the introduction of
a retardation kernel depending on the spatial coordinates only. Such a kernel simulates a kind of non-local interaction of size $\xi$. After deriving an approximated
Fokker-Planck equation we were able to calculate the mean values of the components of the magnetization in the linear spin wave approach. They depend strongly on
the parameters characterizing the retardation (strength $\Gamma_0$, length $\xi$ ) as well as the stochastic (strength $D$, correlation time $\tau_c$) processes.
As a result of the analysis we found that the increase of the retardation strength $\Gamma_0$ compared with the growth of the retardation length $\xi$ can entail
conflictive effects on the lifetime $\tau_L$. The main results are depicted in Fig.~\ref{piclinewidth}. There, in addition to the lifetime of the spin waves
$\tau_L$ the FMR-linewidth $\Delta B$ is displayed. In doing so we want to provide comparability to experimental investigations based on ferromagnetic resonance for cases 
when the LLG is applicable.
Let us remark that also other mechanisms are able to contribute to the damping process. As suggested in \cite{Hillebrands:Rezende:Mills:SDiCFII:2003,Baberschke:PSSB245:2008} 
the Bloch-Bloembergen equations \cite{Bloch:PhysRev70:460:1946,Bloembergen:PhysRev78:572:1950} are more appropriate for in-plane configurations in thin films. 
These equations are characterized by two relaxation times. Another approach with different relaxation processes is based upon the Landau-Lifshitz-Bloch equations 
\cite{Garanin:TheoMatPhys82:169:1990,Garanin:PRB:55:p3050:1997}. Our method including the inevitable stochastic forces can be likewise applied to the modified set of 
equations.\\

One of us (T.B.) is grateful to the Research Network 'Nanostructured Materials'\,,
which is supported by the Saxony-Anhalt State, Germany. Furthermore, we are indebted to Dr. Khalil Zakieri (MPI of Microstructure Physics) for 
valuable discussions.

%

\bibliographystyle{apsrev4-1}
\bibliography{LitNoisyRet}

\begin{thebibliography}{10}%
\makeatletter
\providecommand \@ifxundefined [1]{%
 \ifx #1\undefined \expandafter \@firstoftwo
 \else \expandafter \@secondoftwo
\fi
}%
\providecommand \@ifnum [1]{%
 \ifnum #1\expandafter \@firstoftwo
 \else \expandafter \@secondoftwo
\fi
}%
\providecommand \enquote [1]{``#1''}%
\providecommand \bibnamefont  [1]{#1}%
\providecommand \bibfnamefont [1]{#1}%
\providecommand \citenamefont [1]{#1}%
\providecommand\href[0]{\@sanitize\@href}%
\providecommand\@href[1]{\endgroup\@@startlink{#1}\endgroup\@@href}%
\providecommand\@@href[1]{#1\@@endlink}%
\providecommand \@sanitize [0]{\begingroup\catcode`\&12\catcode`\#12\relax}%
\@ifxundefined \pdfoutput {\@firstoftwo}{%
 \@ifnum{\z@=\pdfoutput}{\@firstoftwo}{\@secondoftwo}%
}{%
 \providecommand\@@startlink[1]{\leavevmode\special{html:<a href="#1">}}%
 \providecommand\@@endlink[0]{\special{html:</a>}}%
}{%
 \providecommand\@@startlink[1]{%
  \leavevmode
  \pdfstartlink
   attr{/Border[0 0 1 ]/H/I/C[0 1 1]}%
   user{/Subtype/Link/A<</Type/Action/S/URI/URI(#1)>>}%
  \relax
 }%
 \providecommand\@@endlink[0]{\pdfendlink}%
}%
\providecommand \url  [0]{\begingroup\@sanitize \@url }%
\providecommand \@url [1]{\endgroup\@href {#1}{\urlprefix}}%
\providecommand \urlprefix [0]{URL }%
\providecommand \Eprint[0]{\href }%
\@ifxundefined \urlstyle {%
  \providecommand \doi [1]{doi:\discretionary{}{}{}#1}%
}{%
  \providecommand \doi [0]{doi:\discretionary{}{}{}\begingroup
  \urlstyle{rm}\Url }%
}%
\providecommand \doibase [0]{http://dx.doi.org/}%
\providecommand \Doi[1]{\href{\doibase#1}}%
\providecommand \bibAnnote [3]{%
  \BibitemShut{#1}%
  \begin{quotation}\noindent
    \textsc{Key:}\ #2\\\textsc{Annotation:}\ #3%
  \end{quotation}%
}%
\providecommand \bibAnnoteFile [2]{%
  \IfFileExists{#2}{\bibAnnote {#1} {#2} {\input{#2}}}{}%
}%
\providecommand \typeout [0]{\immediate \write \m@ne }%
\providecommand \selectlanguage [0]{\@gobble}%
\providecommand \bibinfo [0]{\@secondoftwo}%
\providecommand \bibfield [0]{\@secondoftwo}%
\providecommand \translation [1]{[#1]}%
\providecommand \BibitemOpen[0]{}%
\providecommand \bibitemStop [0]{}%
\providecommand \bibitemNoStop [0]{.\EOS\space}%
\providecommand \EOS [0]{\spacefactor3000\relax}%
\providecommand \BibitemShut [1]{\csname bibitem#1\endcsname}%
\bibitem{HeinrichBland:Book:UMSIIandIII:2005}%
  \BibitemOpen
  \emph{\bibinfo {title} {Ultrathin Magnetic Structures II + III}},\ edited by\
  \bibinfo {editor} {\bibfnamefont{B.}~\bibnamefont{Heinrich}}\ and\ \bibinfo
  {editor} {\bibfnamefont{J.}~\bibnamefont{Bland}}\ (\bibinfo {publisher}
  {Springer},\ \bibinfo {year} {2005})%
  \bibAnnoteFile{NoStop}{HeinrichBland:Book:UMSIIandIII:2005}%
\bibitem{Hillebrands:Rezende:Mills:SDiCFII:2003}%
  \BibitemOpen
  \bibfield{author}{%
  \bibinfo {author} {\bibfnamefont{D.~L.}\ \bibnamefont{Mills}}\ and\ \bibinfo
  {author} {\bibfnamefont{S.~M.}\ \bibnamefont{Rezende}},\ }%
  \enquote{\bibinfo {title} {Spin dynamics in confined magnetic structures ii,
  edited by b. hillebrands and k. ounadjela},}\ \ (\bibinfo {publisher}
  {Springer},\ \bibinfo {address} {Berlin},\ \bibinfo {year} {2003})\ Chap.\
  \bibinfo {chapter} {Spin Damping in Ultrathin Magnetic Films}, pp.\ \bibinfo
  {pages} {27--59}%
  \bibAnnoteFile{NoStop}{Hillebrands:Rezende:Mills:SDiCFII:2003}%
\bibitem{Landau:ZdS:8:p153:1935}%
  \BibitemOpen
  \bibfield{author}{%
  \bibinfo {author} {\bibfnamefont{L.}~\bibnamefont{Landau}}\ and\ \bibinfo
  {author} {\bibfnamefont{E.}~\bibnamefont{Lifshitz}},\ }%
  \bibfield{journal}{%
  \bibinfo {journal} {Zeitschr. d. Sowj.}\ }%
  \textbf{\bibinfo {volume} {8}},\ \bibinfo {pages} {153} (\bibinfo {year}
  {1935})%
  \bibAnnoteFile{NoStop}{Landau:ZdS:8:p153:1935}%
\bibitem{Gilbert:ITOM:40:p3443:2004}%
  \BibitemOpen
  \bibfield{author}{%
  \bibinfo {author} {\bibfnamefont{T.~L.}\ \bibnamefont{Gilbert}},\ }%
  \bibfield{journal}{%
  \bibinfo {journal} {IEEE Trans. Magn.}\ }%
  \textbf{\bibinfo {volume} {40}},\ \bibinfo {pages} {3443} (\bibinfo {year}
  {2004})%
  \bibAnnoteFile{NoStop}{Gilbert:ITOM:40:p3443:2004}%
\bibitem{Malinkowski:APL94:102501:2009}%
  \BibitemOpen
  \bibfield{author}{%
  \bibinfo {author} {\bibfnamefont{G.}~\bibnamefont{Malinowski}}, \bibinfo
  {author} {\bibfnamefont{K.~C.}\ \bibnamefont{Kuiper}}, \bibinfo {author}
  {\bibfnamefont{R.}~\bibnamefont{Lavrijsen}}, \bibinfo {author}
  {\bibfnamefont{H.~J.~M.}\ \bibnamefont{Swagten}},\ and\ \bibinfo {author}
  {\bibfnamefont{B.}~\bibnamefont{Koopmans}},\ }%
  \bibfield{journal}{%
  \bibinfo {journal} {Appl. Phys. Lett.}\ }%
  \textbf{\bibinfo {volume} {94}},\ \bibinfo {pages} {102501} (\bibinfo {year}
  {2009})%
  \bibAnnoteFile{NoStop}{Malinkowski:APL94:102501:2009}%
\bibitem{Gilmore:PhysRevLett.99.027204:2007}%
  \BibitemOpen
  \bibfield{author}{%
  \bibinfo {author} {\bibfnamefont{K.}~\bibnamefont{Gilmore}}, \bibinfo
  {author} {\bibfnamefont{Y.~U.}\ \bibnamefont{Idzerda}},\ and\ \bibinfo
  {author} {\bibfnamefont{M.~D.}\ \bibnamefont{Stiles}},\ }%
  \bibfield{journal}{%
  \bibinfo {journal} {Phys. Rev. Lett.}\ }%
  \textbf{\bibinfo {volume} {99}},\ \bibinfo {pages} {027204} (\bibinfo {year}
  {2007})%
  \bibAnnoteFile{NoStop}{Gilmore:PhysRevLett.99.027204:2007}%
\bibitem{Chubykalo:PhysRevB74:Chubykalo:094436:2006}%
  \BibitemOpen
  \bibfield{author}{%
  \bibinfo {author} {\bibfnamefont{O.}~\bibnamefont{Chubykalo-Fesenko}},
  \bibinfo {author} {\bibfnamefont{U.}~\bibnamefont{Nowak}}, \bibinfo {author}
  {\bibfnamefont{R.~W.}\ \bibnamefont{Chantrell}},\ and\ \bibinfo {author}
  {\bibfnamefont{D.}~\bibnamefont{Garanin}},\ }%
  \bibfield{journal}{%
  \bibinfo {journal} {Phys. Rev. B}\ }%
  \textbf{\bibinfo {volume} {74}},\ \bibinfo {pages} {094436} (\bibinfo {year}
  {2006})%
  \bibAnnoteFile{NoStop}{Chubykalo:PhysRevB74:Chubykalo:094436:2006}%
\bibitem{Garanin:PhysRevB70:212409:2004}%
  \BibitemOpen
  \bibfield{author}{%
  \bibinfo {author} {\bibfnamefont{D.~A.}\ \bibnamefont{Garanin}}\ and\
  \bibinfo {author} {\bibfnamefont{O.}~\bibnamefont{Chubykalo-Fesenko}},\ }%
  \bibfield{journal}{%
  \bibinfo {journal} {Phys. Rev. B}\ }%
  \textbf{\bibinfo {volume} {70}},\ \bibinfo {pages} {212409} (\bibinfo {year}
  {2004})%
  \bibAnnoteFile{NoStop}{Garanin:PhysRevB70:212409:2004}%
\bibitem{Kazantseva:EPL86:27006:2009}%
  \BibitemOpen
  \bibfield{author}{%
  \bibinfo {author} {\bibfnamefont{N.}~\bibnamefont{Kazantseva}}, \bibinfo
  {author} {\bibfnamefont{D.}~\bibnamefont{Hinzke}}, \bibinfo {author}
  {\bibfnamefont{R.~W.}\ \bibnamefont{Chantrell}},\ and\ \bibinfo {author}
  {\bibfnamefont{U.}~\bibnamefont{Nowak}},\ }%
  \bibfield{journal}{%
  \bibinfo {journal} {Europhys. Lett.}\ }%
  \textbf{\bibinfo {volume} {86}},\ \bibinfo {pages} {27006} (\bibinfo {year}
  {2009})%
  \bibAnnoteFile{NoStop}{Kazantseva:EPL86:27006:2009}%
\bibitem{Sparks:PhysRev122:791:1961}%
  \BibitemOpen
  \bibfield{author}{%
  \bibinfo {author} {\bibfnamefont{M.}~\bibnamefont{Sparks}}, \bibinfo {author}
  {\bibfnamefont{R.}~\bibnamefont{Loudon}},\ and\ \bibinfo {author}
  {\bibfnamefont{C.}~\bibnamefont{Kittel}},\ }%
  \bibfield{journal}{%
  \bibinfo {journal} {Phys. Rev.}\ }%
  \textbf{\bibinfo {volume} {122}},\ \bibinfo {pages} {791} (\bibinfo {year}
  {1961})%
  \bibAnnoteFile{NoStop}{Sparks:PhysRev122:791:1961}%
\bibitem{Arias:PhysRevB60:7395:1999}%
  \BibitemOpen
  \bibfield{author}{%
  \bibinfo {author} {\bibfnamefont{R.}~\bibnamefont{Arias}}\ and\ \bibinfo
  {author} {\bibfnamefont{D.~L.}\ \bibnamefont{Mills}},\ }%
  \bibfield{journal}{%
  \bibinfo {journal} {Phys. Rev. B}\ }%
  \textbf{\bibinfo {volume} {60}},\ \bibinfo {pages} {7395} (\bibinfo {year}
  {1999})%
  \bibAnnoteFile{NoStop}{Arias:PhysRevB60:7395:1999}%
\bibitem{Arias:JAP87:5455:2000}%
  \BibitemOpen
  \bibfield{author}{%
  \bibinfo {author} {\bibfnamefont{R.}~\bibnamefont{Arias}}\ and\ \bibinfo
  {author} {\bibfnamefont{D.~L.}\ \bibnamefont{Mills}},\ }%
  \bibfield{journal}{%
  \bibinfo {journal} {J. Appl. Phys.}\ }%
  \textbf{\bibinfo {volume} {87}},\ \bibinfo {pages} {5455} (\bibinfo {year}
  {2000})%
  \bibAnnoteFile{NoStop}{Arias:JAP87:5455:2000}%
\bibitem{Landeros:PhysRevB77:214405:2008}%
  \BibitemOpen
  \bibfield{author}{%
  \bibinfo {author} {\bibfnamefont{P.}~\bibnamefont{Landeros}}, \bibinfo
  {author} {\bibfnamefont{R.~E.}\ \bibnamefont{Arias}},\ and\ \bibinfo {author}
  {\bibfnamefont{D.~L.}\ \bibnamefont{Mills}},\ }%
  \bibfield{journal}{%
  \bibinfo {journal} {Phys. Rev. B}\ }%
  \textbf{\bibinfo {volume} {77}},\ \bibinfo {pages} {214405} (\bibinfo {year}
  {2008})%
  \bibAnnoteFile{NoStop}{Landeros:PhysRevB77:214405:2008}%
\bibitem{Lindner:PhysRevB68:060102:2003}%
  \BibitemOpen
  \bibfield{author}{%
  \bibinfo {author} {\bibfnamefont{J.}~\bibnamefont{Lindner}}, \bibinfo
  {author} {\bibfnamefont{K.}~\bibnamefont{Lenz}}, \bibinfo {author}
  {\bibfnamefont{E.}~\bibnamefont{Kosubek}}, \bibinfo {author}
  {\bibfnamefont{K.}~\bibnamefont{Baberschke}}, \bibinfo {author}
  {\bibfnamefont{D.}~\bibnamefont{Spoddig}}, \bibinfo {author}
  {\bibfnamefont{R.}~\bibnamefont{Meckenstock}}, \bibinfo {author}
  {\bibfnamefont{J.}~\bibnamefont{Pelzl}}, \bibinfo {author}
  {\bibfnamefont{Z.}~\bibnamefont{Frait}},\ and\ \bibinfo {author}
  {\bibfnamefont{D.~L.}\ \bibnamefont{Mills}},\ }%
  \bibfield{journal}{%
  \Doi{10.1103/PhysRevB.68.060102}{\bibinfo {journal} {Phys. Rev. B}}\ }%
  \textbf{\bibinfo {volume} {68}},\ \bibinfo {pages} {060102} (\bibinfo {month}
  {Aug}\ \bibinfo {year} {2003})%
  \bibAnnoteFile{NoStop}{Lindner:PhysRevB68:060102:2003}%
\bibitem{Woltersdorf:PhysRevB69:184417:2004}%
  \BibitemOpen
  \bibfield{author}{%
  \bibinfo {author} {\bibfnamefont{G.}~\bibnamefont{Woltersdorf}}\ and\
  \bibinfo {author} {\bibfnamefont{B.}~\bibnamefont{Heinrich}},\ }%
  \bibfield{journal}{%
  \Doi{10.1103/PhysRevB.69.184417}{\bibinfo {journal} {Phys. Rev. B}}\ }%
  \textbf{\bibinfo {volume} {69}},\ \bibinfo {pages} {184417} (\bibinfo {month}
  {May}\ \bibinfo {year} {2004})%
  \bibAnnoteFile{NoStop}{Woltersdorf:PhysRevB69:184417:2004}%
\bibitem{Lenz:PhysRevB73:144424:2006}%
  \BibitemOpen
  \bibfield{author}{%
  \bibinfo {author} {\bibfnamefont{K.}~\bibnamefont{Lenz}}, \bibinfo {author}
  {\bibfnamefont{H.}~\bibnamefont{Wende}}, \bibinfo {author}
  {\bibfnamefont{W.}~\bibnamefont{Kuch}}, \bibinfo {author}
  {\bibfnamefont{K.}~\bibnamefont{Baberschke}}, \bibinfo {author}
  {\bibfnamefont{K.}~\bibnamefont{Nagy}},\ and\ \bibinfo {author}
  {\bibfnamefont{A.}~\bibnamefont{J\'anossy}},\ }%
  \bibfield{journal}{%
  \bibinfo {journal} {Phys. Rev. B}\ }%
  \textbf{\bibinfo {volume} {73}},\ \bibinfo {pages} {144424} (\bibinfo {year}
  {2006})%
  \bibAnnoteFile{NoStop}{Lenz:PhysRevB73:144424:2006}%
\bibitem{Zakeri:PRB76:2007:104416}%
  \BibitemOpen
  \bibfield{author}{%
  \bibinfo {author} {\bibfnamefont{K.}~\bibnamefont{Zakeri}}, \bibinfo {author}
  {\bibfnamefont{J.}~\bibnamefont{Lindner}}, \bibinfo {author}
  {\bibfnamefont{I.}~\bibnamefont{Barsukov}}, \bibinfo {author}
  {\bibfnamefont{R.}~\bibnamefont{Meckenstock}}, \bibinfo {author}
  {\bibfnamefont{M.}~\bibnamefont{Farle}}, \bibinfo {author}
  {\bibfnamefont{U.}~\bibnamefont{von H\"orsten}}, \bibinfo {author}
  {\bibfnamefont{H.}~\bibnamefont{Wende}}, \bibinfo {author}
  {\bibfnamefont{W.}~\bibnamefont{Keune}}, \bibinfo {author}
  {\bibfnamefont{J.}~\bibnamefont{Rocker}}, \bibinfo {author}
  {\bibfnamefont{S.~S.}\ \bibnamefont{Kalarickal}}, \bibinfo {author}
  {\bibfnamefont{K.}~\bibnamefont{Lenz}}, \bibinfo {author}
  {\bibfnamefont{W.}~\bibnamefont{Kuch}}, \bibinfo {author}
  {\bibfnamefont{K.}~\bibnamefont{Baberschke}},\ and\ \bibinfo {author}
  {\bibfnamefont{Z.}~\bibnamefont{Frait}},\ }%
  \bibfield{journal}{%
  \Doi{10.1103/PhysRevB.76.104416}{\bibinfo {journal} {Phys. Rev. B}}\ }%
  \textbf{\bibinfo {volume} {76}},\ \bibinfo {pages} {104416} (\bibinfo {year}
  {2007})%
  \bibAnnoteFile{NoStop}{Zakeri:PRB76:2007:104416}%
\bibitem{Twisselmann:JAP93:6903:2003}%
  \BibitemOpen
  \bibfield{author}{%
  \bibinfo {author} {\bibfnamefont{D.~J.}\ \bibnamefont{Twisselmann}}\ and\
  \bibinfo {author} {\bibfnamefont{R.~D.}\ \bibnamefont{McMichael}},\ }%
  \bibfield{journal}{%
  \bibinfo {journal} {J. Appl. Phys.}\ }%
  \textbf{\bibinfo {volume} {93}},\ \bibinfo {pages} {6903} (\bibinfo {year}
  {2003})%
  \bibAnnoteFile{NoStop}{Twisselmann:JAP93:6903:2003}%
\bibitem{FaehnleEtAl:JPhysD:0022-3727-41-16-164014}%
  \BibitemOpen
  \bibfield{author}{%
  \bibinfo {author} {\bibfnamefont{M.}~\bibnamefont{F\"ahnle}}, \bibinfo
  {author} {\bibfnamefont{D.}~\bibnamefont{Steiauf}},\ and\ \bibinfo {author}
  {\bibfnamefont{J.}~\bibnamefont{Seib}},\ }%
  \bibfield{journal}{%
  \bibinfo {journal} {J. Phys. D: Applied Physics}\ }%
  \textbf{\bibinfo {volume} {41}},\ \bibinfo {pages} {164014} (\bibinfo {year}
  {2008}),\ \url{http://stacks.iop.org/0022-3727/41/i=16/a=164014}%
  \bibAnnoteFile{NoStop}{FaehnleEtAl:JPhysD:0022-3727-41-16-164014}%
\bibitem{Seib:PhysRevB.79.092418}%
  \BibitemOpen
  \bibfield{author}{%
  \bibinfo {author} {\bibfnamefont{J.}~\bibnamefont{Seib}}, \bibinfo {author}
  {\bibfnamefont{D.}~\bibnamefont{Steiauf}},\ and\ \bibinfo {author}
  {\bibfnamefont{M.}~\bibnamefont{F\"ahnle}},\ }%
  \bibfield{journal}{%
  \Doi{10.1103/PhysRevB.79.092418}{\bibinfo {journal} {Phys. Rev. B}}\ }%
  \textbf{\bibinfo {volume} {79}},\ \bibinfo {pages} {092418} (\bibinfo {year}
  {2009})%
  \bibAnnoteFile{NoStop}{Seib:PhysRevB.79.092418}%
\bibitem{GilmoreStyles:PhysRevB.81.174414}%
  \BibitemOpen
  \bibfield{author}{%
  \bibinfo {author} {\bibfnamefont{K.}~\bibnamefont{Gilmore}}, \bibinfo
  {author} {\bibfnamefont{M.~D.}\ \bibnamefont{Stiles}}, \bibinfo {author}
  {\bibfnamefont{J.}~\bibnamefont{Seib}}, \bibinfo {author}
  {\bibfnamefont{D.}~\bibnamefont{Steiauf}},\ and\ \bibinfo {author}
  {\bibfnamefont{M.}~\bibnamefont{F\"ahnle}},\ }%
  \bibfield{journal}{%
  \Doi{10.1103/PhysRevB.81.174414}{\bibinfo {journal} {Phys. Rev. B}}\ }%
  \textbf{\bibinfo {volume} {81}},\ \bibinfo {pages} {174414} (\bibinfo {year}
  {2010})%
  \bibAnnoteFile{NoStop}{GilmoreStyles:PhysRevB.81.174414}%
\bibitem{BraraasTserkovnyak:PhysRevLett.101.037207}%
  \BibitemOpen
  \bibfield{author}{%
  \bibinfo {author} {\bibfnamefont{A.}~\bibnamefont{Brataas}}, \bibinfo
  {author} {\bibfnamefont{Y.}~\bibnamefont{Tserkovnyak}},\ and\ \bibinfo
  {author} {\bibfnamefont{G.~E.~W.}\ \bibnamefont{Bauer}},\ }%
  \bibfield{journal}{%
  \Doi{10.1103/PhysRevLett.101.037207}{\bibinfo {journal} {Phys. Rev. Lett.}}\
  }%
  \textbf{\bibinfo {volume} {101}},\ \bibinfo {pages} {037207} (\bibinfo {year}
  {2008})%
  \bibAnnoteFile{NoStop}{BraraasTserkovnyak:PhysRevLett.101.037207}%
\bibitem{KardaszHeinrich:PhysRevB.81.094409}%
  \BibitemOpen
  \bibfield{author}{%
  \bibinfo {author} {\bibfnamefont{B.}~\bibnamefont{Kardasz}}\ and\ \bibinfo
  {author} {\bibfnamefont{B.}~\bibnamefont{Heinrich}},\ }%
  \bibfield{journal}{%
  \Doi{10.1103/PhysRevB.81.094409}{\bibinfo {journal} {Phys. Rev. B}}\ }%
  \textbf{\bibinfo {volume} {81}},\ \bibinfo {pages} {094409} (\bibinfo {year}
  {2010})%
  \bibAnnoteFile{NoStop}{KardaszHeinrich:PhysRevB.81.094409}%
\bibitem{LiuMoriyama:PhysRevLett.106.036601}%
  \BibitemOpen
  \bibfield{author}{%
  \bibinfo {author} {\bibfnamefont{L.}~\bibnamefont{Liu}}, \bibinfo {author}
  {\bibfnamefont{T.}~\bibnamefont{Moriyama}}, \bibinfo {author}
  {\bibfnamefont{D.~C.}\ \bibnamefont{Ralph}},\ and\ \bibinfo {author}
  {\bibfnamefont{R.~A.}\ \bibnamefont{Buhrman}},\ }%
  \bibfield{journal}{%
  \Doi{10.1103/PhysRevLett.106.036601}{\bibinfo {journal} {Phys. Rev. Lett.}}\
  }%
  \textbf{\bibinfo {volume} {106}},\ \bibinfo {pages} {036601} (\bibinfo {year}
  {2011})%
  \bibAnnoteFile{NoStop}{LiuMoriyama:PhysRevLett.106.036601}%
\bibitem{WangCui:Nat.Phys.2011:doi10.1038/nphys1928}%
  \BibitemOpen
  \bibfield{author}{%
  \bibinfo {author} {\bibfnamefont{C.}~\bibnamefont{Wang}}, \bibinfo {author}
  {\bibfnamefont{Y.-T.}\ \bibnamefont{Cui}}, \bibinfo {author}
  {\bibfnamefont{J.~A.}\ \bibnamefont{Katine}}, \bibinfo {author}
  {\bibfnamefont{R.~A.}\ \bibnamefont{Buhrman}},\ and\ \bibinfo {author}
  {\bibfnamefont{D.~C.}\ \bibnamefont{Ralph}},\ }%
  \bibfield{journal}{%
  \bibinfo {journal} {Nature Physics (published online 27 Feb.,
  doi:10.1038/nphys1928)}}%
   (\bibinfo {year} {2011}),\ \doi{\bibinfo {doi} {10.1038/nphys1928}}%
  \bibAnnoteFile{NoStop}{WangCui:Nat.Phys.2011:doi10.1038/nphys1928}%
\bibitem{AriasMills:PhysRevB.79.144404}%
  \BibitemOpen
  \bibfield{author}{%
  \bibinfo {author} {\bibfnamefont{R.~E.}\ \bibnamefont{Arias}}\ and\ \bibinfo
  {author} {\bibfnamefont{D.~L.}\ \bibnamefont{Mills}},\ }%
  \bibfield{journal}{%
  \Doi{10.1103/PhysRevB.79.144404}{\bibinfo {journal} {Phys. Rev. B}}\ }%
  \textbf{\bibinfo {volume} {79}},\ \bibinfo {pages} {144404} (\bibinfo {year}
  {2009})%
  \bibAnnoteFile{NoStop}{AriasMills:PhysRevB.79.144404}%
\bibitem{Vilar:RubiPhysRevLett.86.950}%
  \BibitemOpen
  \bibfield{author}{%
  \bibinfo {author} {\bibfnamefont{J.~M.~G.}\ \bibnamefont{Vilar}}\ and\
  \bibinfo {author} {\bibfnamefont{J.~M.}\ \bibnamefont{Rub\'\i{}}},\ }%
  \bibfield{journal}{%
  \Doi{10.1103/PhysRevLett.86.950}{\bibinfo {journal} {Phys. Rev. Lett.}}\ }%
  \textbf{\bibinfo {volume} {86}},\ \bibinfo {pages} {950} (\bibinfo {year}
  {2001})%
  \bibAnnoteFile{NoStop}{Vilar:RubiPhysRevLett.86.950}%
\bibitem{Foros:PRB:79:p214407:2009}%
  \BibitemOpen
  \bibfield{author}{%
  \bibinfo {author} {\bibfnamefont{J.}~\bibnamefont{Foros}}, \bibinfo {author}
  {\bibfnamefont{A.}~\bibnamefont{Brataas}}, \bibinfo {author}
  {\bibfnamefont{G.~E.~W.}\ \bibnamefont{Bauer}},\ and\ \bibinfo {author}
  {\bibfnamefont{Y.}~\bibnamefont{Tserkovnyak}},\ }%
  \bibfield{journal}{%
  \bibinfo {journal} {Phys. Rev. B}\ }%
  \textbf{\bibinfo {volume} {79}},\ \bibinfo {pages} {214407} (\bibinfo {year}
  {2009})%
  \bibAnnoteFile{NoStop}{Foros:PRB:79:p214407:2009}%
\bibitem{Swiebodzinski:PhysRevB.82.144404}%
  \BibitemOpen
  \bibfield{author}{%
  \bibinfo {author} {\bibfnamefont{J.}~\bibnamefont{Swiebodzinski}}, \bibinfo
  {author} {\bibfnamefont{A.}~\bibnamefont{Chudnovskiy}}, \bibinfo {author}
  {\bibfnamefont{T.}~\bibnamefont{Dunn}},\ and\ \bibinfo {author}
  {\bibfnamefont{A.}~\bibnamefont{Kamenev}},\ }%
  \bibfield{journal}{%
  \Doi{10.1103/PhysRevB.82.144404}{\bibinfo {journal} {Phys. Rev. B}}\ }%
  \textbf{\bibinfo {volume} {82}},\ \bibinfo {pages} {144404} (\bibinfo {year}
  {2010})%
  \bibAnnoteFile{NoStop}{Swiebodzinski:PhysRevB.82.144404}%
\bibitem{DiaoNowak:PhysRevLett.104.047202}%
  \BibitemOpen
  \bibfield{author}{%
  \bibinfo {author} {\bibfnamefont{Z.}~\bibnamefont{Diao}}, \bibinfo {author}
  {\bibfnamefont{E.~R.}\ \bibnamefont{Nowak}}, \bibinfo {author}
  {\bibfnamefont{G.}~\bibnamefont{Feng}},\ and\ \bibinfo {author}
  {\bibfnamefont{J.~M.~D.}\ \bibnamefont{Coey}},\ }%
  \bibfield{journal}{%
  \Doi{10.1103/PhysRevLett.104.047202}{\bibinfo {journal} {Phys. Rev. Lett.}}\
  }%
  \textbf{\bibinfo {volume} {104}},\ \bibinfo {pages} {047202} (\bibinfo {year}
  {2010})%
  \bibAnnoteFile{NoStop}{DiaoNowak:PhysRevLett.104.047202}%
\bibitem{HartmannAPL96:2010:082108}%
  \BibitemOpen
  \bibfield{author}{%
  \bibinfo {author} {\bibfnamefont{F.}~\bibnamefont{Hartmann}}, \bibinfo
  {author} {\bibfnamefont{D.}~\bibnamefont{Hartmann}}, \bibinfo {author}
  {\bibfnamefont{P.}~\bibnamefont{Kowalzik}}, \bibinfo {author}
  {\bibfnamefont{L.}~\bibnamefont{Gammaitoni}}, \bibinfo {author}
  {\bibfnamefont{A.}~\bibnamefont{Forchel}},\ and\ \bibinfo {author}
  {\bibfnamefont{L.}~\bibnamefont{Worschech}},\ }%
  \bibfield{journal}{%
  \bibinfo {journal} {Appl. Phys. Lett.}\ }%
  \textbf{\bibinfo {volume} {96}},\ \bibinfo {pages} {082108} (\bibinfo {year}
  {2010})%
  \bibAnnoteFile{NoStop}{HartmannAPL96:2010:082108}%
\bibitem{BoseTrimper:PRB:81:104413:2010}%
  \BibitemOpen
  \bibfield{author}{%
  \bibinfo {author} {\bibfnamefont{T.}~\bibnamefont{Bose}}\ and\ \bibinfo
  {author} {\bibfnamefont{S.}~\bibnamefont{Trimper}},\ }%
  \bibfield{journal}{%
  \bibinfo {journal} {Phys. Rev. B}\ }%
  \textbf{\bibinfo {volume} {81}},\ \bibinfo {pages} {104413} (\bibinfo {year}
  {2010})%
  \bibAnnoteFile{NoStop}{BoseTrimper:PRB:81:104413:2010}%
\bibitem{bosetrimper:retardatin:PRB:2011}%
  \BibitemOpen
  \bibfield{author}{%
  \bibinfo {author} {\bibfnamefont{T.}~\bibnamefont{Bose}}\ and\ \bibinfo
  {author} {\bibfnamefont{S.}~\bibnamefont{Trimper}},\ }%
  \bibfield{journal}{%
  \bibinfo {journal} {submitted to Phys. Rev. B}}%
   (\bibinfo {year} {2011})%
  \bibAnnoteFile{NoStop}{bosetrimper:retardatin:PRB:2011}%
\bibitem{Bar'Yakhtar:DynTopMagSol:Book:1994}%
  \BibitemOpen
  \bibfield{author}{%
  \bibinfo {author} {\bibfnamefont{V.~G.}\ \bibnamefont{Bar'Yakhtar}}, \bibinfo
  {author} {\bibfnamefont{M.~V.}\ \bibnamefont{Chetkin}}, \bibinfo {author}
  {\bibfnamefont{B.~A.}\ \bibnamefont{Ivanov}},\ and\ \bibinfo {author}
  {\bibfnamefont{S.~N.}\ \bibnamefont{Gadetskii}},\ }%
  \emph{\bibinfo {title} {Dynamics of Topological Magnetic Solitons: Experiment
  and Theory (Springer Tracts in Modern Physics)}}\ (\bibinfo {publisher}
  {Springer},\ \bibinfo {year} {1994})%
  \bibAnnoteFile{NoStop}{Bar'Yakhtar:DynTopMagSol:Book:1994}%
\bibitem{Lakshmanan:PA:84:p577:1976}%
  \BibitemOpen
  \bibfield{author}{%
  \bibinfo {author} {\bibfnamefont{M.}~\bibnamefont{Lakshmanan}}, \bibinfo
  {author} {\bibfnamefont{T.~W.}\ \bibnamefont{Ruijgrok}},\ and\ \bibinfo
  {author} {\bibfnamefont{C.~J.}\ \bibnamefont{Thompson}},\ }%
  \bibfield{journal}{%
  \bibinfo {journal} {Physica A}\ }%
  \textbf{\bibinfo {volume} {84}},\ \bibinfo {pages} {577} (\bibinfo {year}
  {1976})%
  \bibAnnoteFile{NoStop}{Lakshmanan:PA:84:p577:1976}%
\bibitem{Kosevich:PRSoPL:194:p117:1990}%
  \BibitemOpen
  \bibfield{author}{%
  \bibinfo {author} {\bibfnamefont{A.~M.}\ \bibnamefont{Kosevich}}, \bibinfo
  {author} {\bibfnamefont{B.~A.}\ \bibnamefont{Ivanov}},\ and\ \bibinfo
  {author} {\bibfnamefont{A.~S.}\ \bibnamefont{Kovalev}},\ }%
  \bibfield{journal}{%
  \bibinfo {journal} {Phys. Rep.}\ }%
  \textbf{\bibinfo {volume} {194}},\ \bibinfo {pages} {117} (\bibinfo {year}
  {1990})%
  \bibAnnoteFile{NoStop}{Kosevich:PRSoPL:194:p117:1990}%
\bibitem{Gardiner:HandbookStochMeth:Book:1990}%
  \BibitemOpen
  \bibfield{author}{%
  \bibinfo {author} {\bibfnamefont{C.~W.}\ \bibnamefont{Gardiner}},\ }%
  \emph{\bibinfo {title} {Handbook of Stochastic Methods for Physics, Chemistry
  and the Natural Sciences}}\ (\bibinfo {publisher} {Springer},\ \bibinfo
  {year} {1990})%
  \bibAnnoteFile{NoStop}{Gardiner:HandbookStochMeth:Book:1990}%
\bibitem{Kampen:StochProcPhysChem:Book:1981}%
  \BibitemOpen
  \bibfield{author}{%
  \bibinfo {author} {\bibfnamefont{N.~G.}\ \bibnamefont{van Kampen}},\ }%
  \emph{\bibinfo {title} {Stochastic Processes in Physics and Chemistry}}\
  (\bibinfo {publisher} {North-Holland},\ \bibinfo {address} {Amsterdam},\
  \bibinfo {year} {1981})%
  \bibAnnoteFile{NoStop}{Kampen:StochProcPhysChem:Book:1981}%
\bibitem{Novikov:SPJ:20:p1290:1965}%
  \BibitemOpen
  \bibfield{author}{%
  \bibinfo {author} {\bibfnamefont{E.~A.}\ \bibnamefont{Novikov}},\ }%
  \bibfield{journal}{%
  \bibinfo {journal} {Sov. Phys. JETP}\ }%
  \textbf{\bibinfo {volume} {20}},\ \bibinfo {pages} {1290} (\bibinfo {year}
  {1965})%
  \bibAnnoteFile{NoStop}{Novikov:SPJ:20:p1290:1965}%
\bibitem{Fox:JoMP:18:p2331:1977}%
  \BibitemOpen
  \bibfield{author}{%
  \bibinfo {author} {\bibfnamefont{R.~F.}\ \bibnamefont{Fox}},\ }%
  \bibfield{journal}{%
  \bibinfo {journal} {J. Math. Phys.}\ }%
  \textbf{\bibinfo {volume} {18}},\ \bibinfo {pages} {2331} (\bibinfo {year}
  {1977})%
  \bibAnnoteFile{NoStop}{Fox:JoMP:18:p2331:1977}%
\bibitem{GarridoSancho:PhysA:1982:479}%
  \BibitemOpen
  \bibfield{author}{%
  \bibinfo {author} {\bibfnamefont{L.}~\bibnamefont{Garrido}}\ and\ \bibinfo
  {author} {\bibfnamefont{J.}~\bibnamefont{Sancho}},\ }%
  \bibfield{journal}{%
  \Doi{DOI: 10.1016/0378-4371(82)90034-6}{\bibinfo {journal} {Physica A}}\ }%
  \textbf{\bibinfo {volume} {115}},\ \bibinfo {pages} {479 } (\bibinfo {year}
  {1982}),\ ISSN \bibinfo {issn} {0378-4371},\
  \url{http://www.sciencedirect.com/science/article/B6TVG-46DF6S6-3R/2/67e78c0%
d74fd3448c16615743fbe26ca}%
  \bibAnnoteFile{NoStop}{GarridoSancho:PhysA:1982:479}%
\bibitem{Dekker:PLA:90:p26:1982}%
  \BibitemOpen
  \bibfield{author}{%
  \bibinfo {author} {\bibfnamefont{H.}~\bibnamefont{Dekker}},\ }%
  \bibfield{journal}{%
  \bibinfo {journal} {Phys. Lett. A}\ }%
  \textbf{\bibinfo {volume} {90}},\ \bibinfo {pages} {26} (\bibinfo {year}
  {1982})%
  \bibAnnoteFile{NoStop}{Dekker:PLA:90:p26:1982}%
\bibitem{Tserkovnyak:PRL:88:p117601:2002}%
  \BibitemOpen
  \bibfield{author}{%
  \bibinfo {author} {\bibfnamefont{Y.}~\bibnamefont{Tserkovnyak}}, \bibinfo
  {author} {\bibfnamefont{A.}~\bibnamefont{Brataas}},\ and\ \bibinfo {author}
  {\bibfnamefont{G.~E.~W.}\ \bibnamefont{Bauer}},\ }%
  \bibfield{journal}{%
  \bibinfo {journal} {Phys. Rev. Lett.}\ }%
  \textbf{\bibinfo {volume} {88}},\ \bibinfo {pages} {117601} (\bibinfo {year}
  {2002})%
  \bibAnnoteFile{NoStop}{Tserkovnyak:PRL:88:p117601:2002}%
\bibitem{Katine:PhysRevLett:84:3149}%
  \BibitemOpen
  \bibfield{author}{%
  \bibinfo {author} {\bibfnamefont{J.~A.}\ \bibnamefont{Katine}}, \bibinfo
  {author} {\bibfnamefont{F.~J.}\ \bibnamefont{Albert}}, \bibinfo {author}
  {\bibfnamefont{R.~A.}\ \bibnamefont{Buhrman}}, \bibinfo {author}
  {\bibfnamefont{E.~B.}\ \bibnamefont{Myers}},\ and\ \bibinfo {author}
  {\bibfnamefont{D.~C.}\ \bibnamefont{Ralph}},\ }%
  \bibfield{journal}{%
  \Doi{10.1103/PhysRevLett.84.3149}{\bibinfo {journal} {Phys. Rev. Lett.}}\ }%
  \textbf{\bibinfo {volume} {84}},\ \bibinfo {pages} {3149} (\bibinfo {year}
  {2000})%
  \bibAnnoteFile{NoStop}{Katine:PhysRevLett:84:3149}%
\bibitem{Landau:StatPhysPart2:Book:1980}%
  \BibitemOpen
  \bibfield{author}{%
  \bibinfo {author} {\bibfnamefont{L.~D.}\ \bibnamefont{Landau}}, \bibinfo
  {author} {\bibfnamefont{E.}~\bibnamefont{Lifshitz}},\ and\ \bibinfo {author}
  {\bibfnamefont{L.}~\bibnamefont{Pitaevskii}},\ }%
  \emph{\bibinfo {title} {Statistical Physics Part 2: Theory of the Condensed
  State}}\ (\bibinfo {publisher} {Pergamon Press},\ \bibinfo {address}
  {Oxford},\ \bibinfo {year} {1980})%
  \bibAnnoteFile{NoStop}{Landau:StatPhysPart2:Book:1980}%
\bibitem{Baberschke:PSSB245:2008}%
  \BibitemOpen
  \bibfield{author}{%
  \bibinfo {author} {\bibfnamefont{K.}~\bibnamefont{Baberschke}},\ }%
  \bibfield{journal}{%
  \bibinfo {journal} {phys. stat. sol. (b)}\ }%
  \textbf{\bibinfo {volume} {245}},\ \bibinfo {pages} {174} (\bibinfo {year}
  {2008})%
  \bibAnnoteFile{NoStop}{Baberschke:PSSB245:2008}%
\bibitem{Bloch:PhysRev70:460:1946}%
  \BibitemOpen
  \bibfield{author}{%
  \bibinfo {author} {\bibfnamefont{F.}~\bibnamefont{Bloch}},\ }%
  \bibfield{journal}{%
  \bibinfo {journal} {Phys. Rev.}\ }%
  \textbf{\bibinfo {volume} {70}},\ \bibinfo {pages} {460} (\bibinfo {year}
  {1946})%
  \bibAnnoteFile{NoStop}{Bloch:PhysRev70:460:1946}%
\bibitem{Bloembergen:PhysRev78:572:1950}%
  \BibitemOpen
  \bibfield{author}{%
  \bibinfo {author} {\bibfnamefont{N.}~\bibnamefont{Bloembergen}},\ }%
  \bibfield{journal}{%
  \bibinfo {journal} {Phys. Rev.}\ }%
  \textbf{\bibinfo {volume} {78}},\ \bibinfo {pages} {572} (\bibinfo {year}
  {1950})%
  \bibAnnoteFile{NoStop}{Bloembergen:PhysRev78:572:1950}%
\bibitem{Garanin:TheoMatPhys82:169:1990}%
  \BibitemOpen
  \bibfield{author}{%
  \bibinfo {author} {\bibfnamefont{D.~A.}\ \bibnamefont{Garanin}}, \bibinfo
  {author} {\bibfnamefont{V.~V.}\ \bibnamefont{Ishchenko}},\ and\ \bibinfo
  {author} {\bibfnamefont{L.~V.}\ \bibnamefont{Panina}},\ }%
  \bibfield{journal}{%
  \bibinfo {journal} {Theor. Math. Phys.}\ }%
  \textbf{\bibinfo {volume} {82}},\ \bibinfo {pages} {169} (\bibinfo {year}
  {1990})%
  \bibAnnoteFile{NoStop}{Garanin:TheoMatPhys82:169:1990}%
\bibitem{Garanin:PRB:55:p3050:1997}%
  \BibitemOpen
  \bibfield{author}{%
  \bibinfo {author} {\bibfnamefont{D.~A.}\ \bibnamefont{Garanin}},\ }%
  \bibfield{journal}{%
  \bibinfo {journal} {Phys. Rev. B}\ }%
  \textbf{\bibinfo {volume} {55}},\ \bibinfo {pages} {3050} (\bibinfo {year}
  {1997})%
  \bibAnnoteFile{NoStop}{Garanin:PRB:55:p3050:1997}%
\end{thebibliography}%

\end{document}